\documentclass[journal,nofonttune]{IEEEtran}
\pdfoutput=1

\IEEEoverridecommandlockouts
\usepackage{cite}
\usepackage{amsmath,amssymb,amsfonts,amsthm}
\usepackage{commath} 

\usepackage{algorithm}
\usepackage{algpseudocode}

\usepackage{graphicx}
\usepackage{textcomp}
\usepackage{lipsum} 
\usepackage{xcolor}
\def\BibTeX{{\rm B\kern-.05em{\sc i\kern-.025em b}\kern-.08em
    T\kern-.1667em\lower.7ex\hbox{E}\kern-.125emX}}
\usepackage{float}
\usepackage{caption}
\usepackage{subcaption}
\usepackage[nolist]{acronym}
\usepackage[permil]{overpic}
\usepackage{hyperref}

\newcommand{\iu}{\textrm{j}}
\newcommand{\vect}[1]{\boldsymbol{\mathbf{#1}}}

\newcommand{\hide}[1]{}

\newtheorem{lemma}{Lemma}

\def\Htran{\mathsf{H}}
\def\Ttran{\mathsf{T}}

\allowdisplaybreaks

\title{{A Novel Near-Field Dictionary Design for Hybrid MIMO with Uniform Planar Arrays}}

\author{L. Antonelli,
        Antonio A.~D'Amico,~\IEEEmembership{Senior~Member,~IEEE,}
        and~Luca~Sanguinetti,~\IEEEmembership{Fellow,~IEEE}\vspace{-0.8cm}
\thanks{ \indent An earlier version of this paper was presented at IEEE Asilomar 2025 \cite{antonelli2026enhancedpolardomaindictionarydesign}.

\indent The authors are with the Dipartimento di Ingegneria dell’Informazione, University of Pisa, 56126 Pisa, Italy (e-mail: luca.antonelli@phd.unipi.it; antonio.damico@unipi.it;
luca.sanguinetti@unipi.it).

\indent This work was supported in part by the framework of the HORIZON-JU-SNS-2022 project TIMES, cofunded by the European Union, in part by the Italian Ministry of Education and Research (MUR) in the framework of the FoReLab Project (Department of Excellence), and in part
by the project GARDEN, funded by EU in NextGenerationEU Plan through Italian “Bando Prin 2022-D.D.1409 del 14-09-2022".}}

\begin{document}

\begin{acronym}
    \acro{THz}{TeraHertz}
    \acro{GHz}{GigaHertz}
    \acro{MHz}{MegaHertz}
    \acro{UE}{user equipment}
    \acro{i.i.d.}{independent-and-identically-distributed}
    \acro{SNR}{signal-to-noise ratio}
     \acro{NMSE}{Normalized Mean Square estimation
    Error}
    \acro{RF}{radio frequency}
    \acro{U-MIMO}{Ultra-massive MIMO}
    \acro{MIMO}{Multiple Output Multiple Input}
    \acro{BS}{base station}
    \acro{ULA}{uniform linear array}
    \acro{UPA}{uniform planar array}
    \acro{URA}{uniform rectangular array}
    \acro{UCA}{uniform circular array}
    \acro{ERD}{effective-Rayleigh distance}
    \acro{FC}{fully-connected}
    \acro{P-SOMP}{polar simultaneous orthogonal matching pursuit}
    \acro{FDP}{Fully Digital P-SOMP}
    \acro{P-SIGW}{polar simultaneous gridless weighted}
    \acro{SE}{spectral efficiency}
    \acro{CSI}{channel state information}
    \acro{RoI}{region-of-interest}
    \acro{DoF}{degree-of-freedom}
    \acro{RP}{reference plane}
    \acro{MR}{maximum ratio}
    \acro{MMSE}{minimum mean square error}
    \acro{LOS}{line-of-sight}
    \acro{NLOS}{non-line-of-sight}
    \acro{LS}{least squares}
    \acro{CS}{compressed sensing}
    \acro{OMP}{orthogonal matching pursuit}
    \acro{SOMP}{simultaneous-OMP}
    \acro{BP}{basis pursuit}
    \acro{LASSO}{least absolute shrinkage and selection operator}
    \acro{mmWave}{millimeter-wave}
    \acro{DFT}{discrete Fourier transform}
\end{acronym}

\acresetall

\maketitle
\begin{abstract}
   Near-field ultra-massive MIMO (U-MIMO) systems provide enhanced spatial resolution but present challenges for channel estimation, particularly when hybrid architectures are employed. Within this framework, dictionary-based channel estimation schemes are needed to achieve accurate reconstruction from a reduced set of measurements. However, existing near-field dictionaries generally provide full three-dimensional coverage, which is unnecessary when user equipments are primarily located on the ground. In this paper, we propose a novel near-field grid design tailored to this common scenario. Specifically, grid points lie on a reference plane located at an arbitrary height with respect to the U-MIMO system, equipped with a uniform planar array. Furthermore, a channel accuracy metric is used to improve codebook performance, and to remark the limitations of the traditional far-field angular sampling in the near field. Results show that, as long as user equipments are not far from the reference plane, the proposed grid outperforms state-of-the-art designs in both channel estimation accuracy and spectral efficiency.
\end{abstract}

\begin{IEEEkeywords}
    Ultra-massive MIMO, near field, far field, channel estimation, spectral efficiency, hybrid architecture.\vspace{-0.4cm}
\end{IEEEkeywords}

\section{Introduction}
\ac{U-MIMO} is envisioned as a key technology to meet the high traffic demands of future wireless communication systems \cite{9166263,5764977,rappaport2019wireless}. By exploiting the short wavelengths at \ac{mmWave} or \ac{THz} frequencies, \ac{U-MIMO} systems densely pack many antennas into a compact area, enabling substantial gains in beam focusing and spatial multiplexing \cite{bjornson2019massive}. However, the conventional fully-digital architecture, where each antenna is connected to a dedicated \ac{RF} chain, becomes prohibitively power-intensive for \ac{U-MIMO} systems \cite{ning2023beamforming}. Consequently, hybrid architectures, which limit the number of \ac{RF} chains \cite{9508929}, are typically adopted instead.

Unlike sub-$6$ GHz communications, where \acp{UE} are typically located in the far field, 
the very large number of antennas operating at short wavelengths in \ac{U-MIMO} systems, generally locates \acp{UE} in the radiating near field \cite{bacci2023spherical}. This shift requires the development of near-field dictionaries, where, differently from far-field scenarios, the range domain is crucial due to the limited beamforming depth \cite{bjornson2021primer,ramezani2023near,cui2022near}.

\subsection{Literature review}

Since near-field codebooks are built sampling both the angular and range domains, they are typically large. However, several \ac{CS} algorithms whose overheads are independent of the codebook's size, have been used for channel estimation with low pilot overhead. For instance, \cite{lee2016channel} exploits the angular-domain sparsity to estimate the channel using the classical \ac{OMP} algorithm. Moreover, \cite{rodriguez2018frequency} proposes a variant of the \ac{OMP}, referred to as \ac{SOMP}, which handles the colored noise introduced by hybrid architectures.

Notably, all above solutions exploit the channel sparsity in the angular domain, which is available in the far field as the wavefront is approximately planar. However, in the near field, as the curvature of the wavefront is not negligible, the energy spreads into multiple angles \cite{cui2022channel}. Consequently, the channel is significantly less sparse in the angular domain, which causes the traditional approaches based on the classical far-field dictionary to suffer a severe performance degradation. Accordingly, novel dictionary designs are needed to make the channel sparse in the near field. Recently, a polar-domain near-field design has been proposed in \cite{cui2022channel} for a \ac{ULA}. Here, the design is based on the dictionary column coherence, which is more challenging to control in the near field.

Since \acp{UPA} allow to pack more antennas within a limited area, the polar-domain dictionary in \cite{cui2022channel} has been extended to the near field of a uniform rectangular array in \cite{wu2023multiple}. Furthermore, \cite{demir2023new} generalizes this design to arbitrary antenna spacings and enables improved control of the column coherence. Additionally, a concentric-ring dictionary has been proposed in \cite{wu2023enabling} for a uniform circular array, whose symmetry is leveraged to achieve uniform and extended near-field regions across all angular directions, thereby enabling more \acp{UE} to benefit from near-field beamforming. The boundary for effective near-field beamforming is investigated in \cite{hussain2024near}, leading to the development of a novel polar codebook, tailored to the concept of \ac{ERD}. Moreover, the combined use of the far-field and polar dictionaries has been recently proposed. In \cite{zhang2022fast}, a \ac{DFT}-based codebook is first employed to estimate the \ac{UE} direction, and a polar codebook is subsequently used to estimate the corresponding distance. Following a similar approach, \cite{wu2023two} splits the array into two subarrays: one is associated with a far-field dictionary for angular beam training, the other with a polar dictionary for range estimation.

State-of-the-art dictionaries for \acp{UPA}, such as \cite{wu2023multiple,demir2023new,wu2023enabling,hussain2024near,zhang2022fast,wu2023two}, typically provide full spatial coverage, meaning that they are designed for arbitrary \ac{UE} locations in the three-dimensional space. However, this is unnecessary when \acp{UE} are distributed close to a planar surface, which commonly occurs in practice.

\subsection{Contributions}
 Our main contributions can be summarized as follows.
\begin{itemize}
    \item First, we show that when the number of observations exceeds the number of antennas, a dictionary-based approach is unnecessary, as the LS estimator is generally accurate -- especially in the near field, where the SNR is higher. However, with few observations, LS becomes unreliable, while sparse recovery methods such as P-SOMP can still yield accurate channel estimates by exploiting sparsity in a suitable dictionary domain. Since near-field channels are not sparse in the traditional far-field codebook domain, near-field dictionaries are required.
    \item Since state-of-the-art near-field dictionaries for \ac{UPA} typically provide full three-dimensional coverage, we propose a novel near-field grid, specifically tailored to \acp{UE} distributed on a \ac{RP}, rather than throughout the entire three-dimensional space. To further assess its practical applicability, we evaluate the robustness of the proposed design under mismatch conditions in which \acp{UE} are located within a three-dimensional volume, as frequently occurs in practice. Results show that, as long as \acp{UE} remain sufficiently close to the \ac{RP} (within a few meters), the proposed design outperforms state-of-the-art dictionaries designed for full three-dimensional coverage.
    \item Although recent dictionary designs - such as those in \cite{cui2022channel,demir2023new} - are based on the codebook column coherence, such metric is not directly indicative of system performance. To address this, we introduce the optimal NMSE $(\rm NMSE_{opt})$, which better reflects codebook performance. As discussed in Sec.~\ref{sec:grid_construction}, minimizing $\rm NMSE_{opt}$ often yields angularly over-sampled grids relative to the traditional far-field angular sampling -- commonly adopted even in near-field dictionaries -- which calls into question its applicability for near-field applications.
\end{itemize}

\subsection{Organization and Notation}

The remainder of this paper is organized as follows. Section II provides the system and signal model and describes the \ac{LS} and \ac{P-SOMP} algorithms used for channel estimation. In Section III, the far-field dictionary performance is investigated, particularly in the near field. In Section IV, the proposed novel near-field grid is described. Simulations are carried out in Section V, and conclusions are drawn in Section VI.   

Lower-case and upper-case boldface letters represent vectors and matrices, respectively; $X_{i,j}$ denotes the element on row $i$ and column $j$ of the matrix $\vect{X}$; $X_i$ denotes the row $i$ of the matrix $\vect{X}$. $(\cdot)^\Ttran$, $(\cdot)^*$, $(\cdot)^H$, $(\cdot)^\dagger$ indicate the transpose, the conjugate, the Hermitian, and the Moore-Penrose inverse, respectively; $|\cdot|$, $||\cdot||$ denote the absolute operator and the Euclidean norm, respectively; $\mathcal{CN}(\mu, \Sigma)$ and $\mathcal{U}(a, b)$ represent the complex Gaussian distribution
with mean $\mu$ and covariance $\Sigma$, and the uniform distribution between $a$ and $b$, respectively; mod$(c,d)$ denotes the remainder of $c$ divided by $d$; $\lfloor \cdot \rfloor$ indicates the floor function; $\otimes$ denotes the Kronecker product.
 
\section{System Model}
\begin{figure}[t]
    \centering
    \includegraphics[width=0.9\columnwidth]{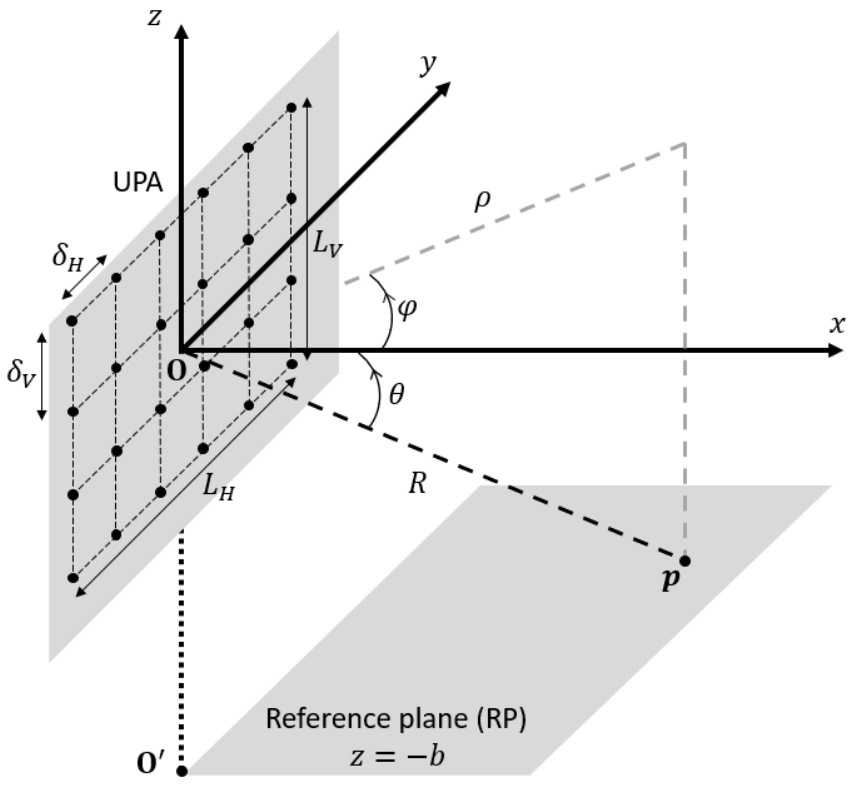}
    \caption{Pictorial illustration of our system model.\vspace{-0.3cm}}
    \label{fig:1}
\end{figure}

We consider the uplink of a U-MIMO system with $K$ single-antenna active \acp{UE}. The carrier frequency is $f_c$, with corresponding wavelength $\lambda$, and the bandwidth is indicated by $B$.

The \ac{BS} deploys a hybrid architecture with $N_{RF} \geq K$ \ac{RF} chains in a \textit{fully-connected configuration}, where each antenna is connected to every \ac{RF} chain. Moreover, the \ac{BS} is equipped with the \ac{UPA} shown in Fig.~\ref{fig:1}. The array lies on the $YZ$ plane, with the $Y$-axis and the $Z$-axis defining the horizontal and vertical directions, respectively. 
We denote with $M_H$ and $M_V$ the numbers of horizontal and vertical antennas, respectively, and indicate with $\delta_H$ and $\delta_V$ their corresponding spacings. Accordingly, the total number of antennas is $M = M_H M_V$, which we assume to be much larger than the number of \ac{RF} chains, i.e., $N_{\rm RF} \ll M$. The horizontal and vertical apertures are given by $L_H = (M_H - 1)\delta_H$ and $L_V = (M_V - 1)\delta_V$, respectively, and the total array aperture is $L = \sqrt{L_H^2+L_V^2}$.
The origin of the Cartesian reference system $\vect{O}$ coincides with the center of the array, which is placed at a height offset $b$ with respect to a \ac{RP} where \acp{UE} are displaced. Antennas are numbered from left to right and from bottom to top, so that the coordinates $\mathbf{u}_m = (x_m,y_m,z_m)$ of the $m$th antenna, with $m = 1, 2,\ldots,M$, are 
\begin{align} \label{eq: AntennaCoordinates}
    x_m =0 \quad y_m =i(m) \delta_H \quad \quad z_m =j(m) \delta_V
\end{align}
with 
\begin{align}
    i(m) &= \mod\left(m-1,M_H\right) - \dfrac{M_H-1}{2} \\
    j(m) &= \left\lfloor \dfrac{m-1}{M_H} \right \rfloor - \dfrac{M_V-1}{2}
\end{align}
where we have implicitly assumed that $M_H$ and $M_V$ are odd.

The \ac{LOS} channel from each \ac{UE} to the array is given by $\vect{h} = \left[h_1,h_2, \dots h_M\right]$ with
\begin{equation} \label{eq:multipath_spherical_channel}
    h_{m} = \sqrt{\beta_{m}}e^{-j\frac{2\pi}{\lambda}r_{m}}
\end{equation}
where $r_m$ is the distance between the \ac{UE} and the antenna $m$, and $\beta_{m}$ accounts for the free space path loss.

\subsection{Channel estimation}
We consider a standard time–frequency division duplex protocol \cite{massivemimobook}, where $\tau_c$ channel uses are available for uplink channel estimation and data transmission. The $K$ \acp{UE} transmit simultaneously over $K$ subcarriers using orthogonal frequency-domain pilot sequences across $\tau$ time slots. Thus, channel estimation requires $\tau_p = K\tau$ resources, and the remaining $\tau_c - \tau_p$ are allocated to data transmission. Moreover, we assume that each \ac{UE} moves with speed $v$ and compute the coherence time according to the rule-of-thumb $T_c = \frac{\lambda}{10v}$ \cite{massivemimobook}. Since we consider a \ac{LOS} scenario, where the coherence bandwidth is generally large, we assume that it coincides with the bandwidth $B$. Hence, the coherence block size is $\tau_c = T_cB$.

Signals at different antennas are combined by means of analogue combiners $\vect{a}_{i,r}\in \mathbb{C}^M$ for $i=1,\ldots,\tau$ and $r=1,\ldots,N_{\rm{RF}}$, whose entries are randomly selected in $\{\pm 1/\sqrt{M}\}$ with equal probabilities. The notation indicates that different combiners may be used in different time slots and different \ac{RF} chains.  We denote as $\vect{A}_i = [\vect{a}_{i,1},\ldots, \vect{a}_{i,N_{\rm{RF}}}] \in \mathbb{C} ^{M \times N_{\rm{RF}}}$ the analog combining matrix for the slot $i$, and with $\vect{A} = [\mathbf{A}_1 \, \cdots \mathbf{A}_{\tau}]^{\Ttran} \in \mathbb{C}^{N_{\rm{RF}}\tau \times M}$ the overall combining matrix.

Assuming perfect orthogonality of the received pilot sequences, channel estimation can be performed independently for each \ac{UE}. Accordingly, we drop the user-index $k$ and focus on a generic user. After straightforward calculations, the overall pilot signal during the channel estimation phase\hide{for any \ac{UE}} is given by
\begin{equation} \label{eq:pilot_vector}
    \vect{y} = \sqrt{p} K \vect{A} \vect{h} + \vect{n}
\end{equation}
where $p$ is the transmit power (assumed identical for all UEs), and $\vect{n} = [\vect{n}^{\Ttran}_1  \, \cdots \, \vect{n}^{\Ttran}_{\tau} ]^{\Ttran} \in \mathbb{C}^{N_{\rm{RF}}\tau}$ with $\vect{n}_{i} \sim \mathcal{CN}(\vect{0},\sigma^2 K \vect{A}^{\mathsf{T}}_{i}\vect{A}_{i}^*)$ collects the noise. We assume that $\sigma^2 = N_0B$, with $N_0$ denoting the noise power spectral density.
The noise correlation matrix is given by $\vect{C} = \rm blkdiag(\vect{C}_1\vect{C}_1^H, \dots , \vect{C}_{\tau}\vect{C}_{\tau}^H)$, where $\vect{C}_i = \sigma^2K\vect{A}_i^\Ttran\vect{A}_i^*$. 
Accordingly, the whitened pilot signal is
\begin{equation} \label{eq:pilot_whitened_signal}
    \vect{z} = \vect{L^{-1}}\vect{y} = \sqrt{p}K\vect{L}^{-1}\vect{Ah} + \vect{w}
\end{equation}
where $\vect{w} \sim \mathcal{CN}(\vect{0},\vect{I}_{ N_{\rm{RF}}\tau})$ and the whitening matrix $\vect{L}$ is obtained from the Cholesky decomposition of $\vect{C}$, i.e., $\vect{C} = \vect{L}\vect{L}^H$. From \eqref{eq:pilot_vector}, the channel can be estimated using the \ac{LS} estimator
\begin{align}
\hat{\vect{h}}_{\rm LS} = \frac{\left(\vect{L^{-1}A}\right)^\dagger}{\sqrt{p}K}\vect{z}.
\end{align}
When the number of observations is smaller than the number of antennas, i.e., $N_{\rm{RF}}\tau < M$, the \ac{LS} becomes inaccurate \cite{berger2010application}. However, if the channel admits a sparse representation in an appropriate dictionary domain, reliable estimation is still possible using alternative algorithms \cite{donoho2006compressed}.

\subsection{Dictionary-based channel estimation}
Let us consider an arbitrary dictionary $\vect{W} \in \mathbb{C}^{M \times Q}$, and assume that rank$(\vect{W}) = M$. By doing so, any channel $\vect{h} \in \mathbb{C}^{M}$ can be represented in terms of its projection $\vect{h}^\mathcal{P} \in \mathbb{C}^{Q}$ into the dictionary domain as\footnote{Since rank$(\vect{W}) = M$, we have implicitly assumed that $Q \geq M$. In this paper, we focus on the case $Q = M$, which means that $\vect{W}$ is a basis of $\mathbb{C}^{M}$, and the projection $\vect{h}^\mathcal{P}$ is therefore unique.}
\begin{equation} \label{eq:channel_projection}
    \vect{h} = \vect{W}\vect{h}^\mathcal{P}. 
\end{equation}
Substituting \eqref{eq:channel_projection} into~\eqref{eq:pilot_whitened_signal}, the whitened pilot signal becomes
\begin{align}
    \vect{z} = \sqrt{p}K\vect{L}^{-1}\vect{A}\vect{W}\vect{h}^\mathcal{P} + \vect{w} = \vect{\Psi}\vect{h}^\mathcal{P} + \vect{w} \label{eq:pilot_whitened_signal_W}
\end{align}
where we have defined 
\begin{equation} \label{eq:Psi_dictionary}
    \vect{\Psi} =\sqrt{p}K\vect{L}^{-1}\vect{A}\vect{W} \in \mathbb{C}^{N_{\rm{RF}}\tau\times Q}.
\end{equation}
It is well known from \ac{CS} theory \cite{donoho2006compressed} that, when $\vect{h}^\mathcal{P}$ is sparse — i.e., it
contains only a few significant entries — estimating $\vect{h}$ directly is suboptimal. Instead, a more effective approach is to first estimate the channel in the dictionary domain \cite{donoho2006compressed}. Specifically, an estimate $\widehat{\vect{h}}^\mathcal{P}$ is obtained and subsequently mapped to the physical channel via \eqref{eq:channel_projection}.

To obtain sparse channel representations, dictionary design plays a fundamental role. In Sec.~\ref{sec:far_field_review}, we first review the conventional far-field dictionary and show that it does not provide high sparsity, particularly in the near field. This limitation has motivated the development of several state-of-the-art near-field dictionaries, which are typically constructed as collections of steering vectors evaluated over suitable spatial grids. When the \ac{BS} deploys a planar array, these design typically provide full three-dimensional spatial coverage. However, such generality is often unnecessary when \acp{UE} are near a planar surface, which can commonly arise in practice. Therefore, in Sec.~\ref{sec:proposed_design} we propose a novel near-field grid design tailored to this scenario.

Once sparsity has been induced through an appropriate dictionary, dedicated algorithms are required to effectively exploit it. Since the dictionary size $Q$ generally exceeds the number of antennas $M$, the dictionary is not a basis of $\mathbb{C}^M$. Consequently, the representation $\vect{h}^\mathcal{P} \in \mathbb{C}^Q$ of the channel $\vect{h} \in \mathbb{C}^M$ in the dictionary domain is not unique. Among all possible representations, we seek for the sparse ones, which must be estimated through suitable sparse-recovery algorithms. To this end, several schemes have been proposed, including Basis Pursuit (BP) \cite{chen2001atomic} and the \ac{LASSO} \cite{elad2010sparse}. Among these, we focus on the \ac{P-SOMP} algorithm \cite{tropp2006algorithms}.

\section{Review of far-field dictionary design} \label{sec:far_field_review}
\begin{figure}[t]
    \centering\vspace{-0.3cm}
    \includegraphics[width=\columnwidth]{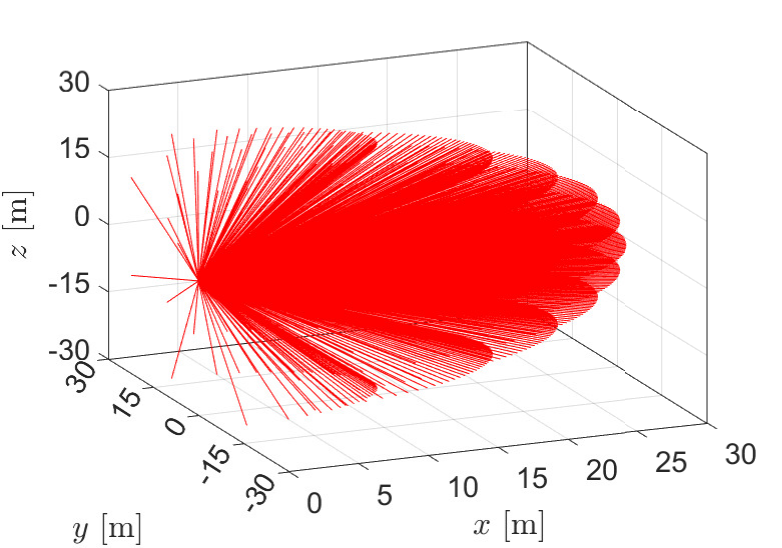}
        \caption{Traditional far-field dictionary directions with $\lambda$ = $1$\,cm, $M_H = 101$, $M_V = 11$ and $\delta_H = \delta_V = \lambda/2$.\vspace{-0.3cm}}
        \label{fig:2}
\end{figure}

When the \ac{BS} deploys an \ac{UPA}, the traditional far-field dictionary coincides with the $2D$-Fourier basis given by
\begin{align} \label{eq:2D_fourier_basis}
    \vect{W}^{\rm{ff}} = \left[\vect{\phi}_{0,0},\ \dots \, \vect{\phi}_{u,v}, \  \dots \ ,\vect{\phi}_{M_H-1,M_V-1}\right]^\Ttran \in \mathbb{C}^{M \times M}
\end{align}
where $\vect{\phi}_{u,v} = \vect{\phi}_u(x) \otimes \vect{\phi}_v(y) \in \mathbb{C}^{M}$ with
\begin{align} \label{eq:fourier_basis_vector}
    \vect{\phi}_u(x)= \frac{1}{\sqrt{M_H}}e^{j2\pi\frac{ux}{M_H}} \quad x \in \{0,1, \ \dots \ , M_H-1\} \\
     \vect{\phi}_v(y)= \frac{1}{\sqrt{M_V}}e^{j2\pi\frac{vy}{M_V}} \quad y \in \{0,1, \ \dots\ ,M_V-1\}
\end{align}
for $u \in \{0,1, \dots ,M_H-1\}$ and $v \in \{0,1, \dots, M_V-1\}$.
The $2$D-Fourier basis in \eqref{eq:2D_fourier_basis} can be interpreted as a collection of far-field steering vectors evaluated along specific spatial directions \cite{balanis2016antenna}. Based on \cite[Sec. 2.2]{naidu2009sensor}, the direction\footnote{When either $\delta_H \geq \lambda/2$ or $\delta_V \geq \lambda/2$, multiple pairs $(U,V)$ may be compatible with \eqref{eq:theta_far_field} and~\eqref{eq:phi_far_field}. In this case, some basis elements $\vect{\phi}_{u,v}$ coincide with far-field steering vectors corresponding to distinct directions, a phenomenon commonly referred to as spatial aliasing.} $(\varphi,\theta)$ for the arbitrary basis element $\vect{\phi}_{u,v}$ can be written as 
\begin{align}
    \theta &= \arcsin{\left(\frac{V\lambda}{M_V\delta_V}\right)} \label{eq:theta_far_field} \\
    \varphi &= \arcsin{\left(\frac{U\lambda}{\cos{\theta}M_H\delta_H}\right)}\label{eq:phi_far_field}
\end{align}
where $U,V$ are integer numbers chosen compatibly with $u = \text{mod}(U,M_H)$ and $v = \text{mod}(V,M_V)$. As explained next, we use \eqref{eq:theta_far_field} and~\eqref{eq:phi_far_field} to distinguish between two different scenarios, referred to as \textit{on-dictionary} and \textit{off-dictionary}. Fig.~\ref{fig:2} shows the directions covered by $2D$-Fourier basis for $\lambda$ = $1$\,cm, $M_H = 101$, $M_V = 11$ and $\delta_H = \delta_V = \lambda / 2$.

\subsection{Channel Estimation Accuracy}
Now, we evaluate the performance of the far-field dictionary in terms of NMSE of channel estimates. 
For channel estimation, we use the \ac{P-SOMP} algorithm with the far-field dictionary, i.e., $\vect{W} = \vect{W}^{\rm{ff}}$, and set the sparsity order to $N^\mathcal{P} = 1$, which we assume throughout the paper. For comparison, we also report the performance of the \ac{LS} estimator. We consider a single-user scenario, i.e., $K = 1$, and set its transmit power to $p = 15$\,dBm. We assume that the \ac{UE} has random coordinates 
\begin{equation}
    (R\cos{\theta}\cos{\varphi},R\cos{\theta}\sin{\varphi},R\sin{\theta})
\end{equation}
with $R \sim \mathcal{U}(20,100)$\,m, and consider two scenarios:
\begin{itemize}
    \item \textit{On-dictionary}: The \ac{UE} is located only along directions covered by the far-field dictionary $\vect{W}^{\rm{ff}}$, i.e., the direction $(\varphi,\theta)$ is chosen compatibly with \eqref{eq:theta_far_field} and~\eqref{eq:phi_far_field}.
    \item \textit{Off-dictionary}: The \ac{UE} is distributed along a random direction, i.e., $\varphi \sim \mathcal{U}(-\pi/2,\pi/2)$ and $\theta \sim \mathcal{U}(-\pi/2,\pi/2)$.
\end{itemize}
All other simulation parameters are given in Tab.~\ref{tab:system_parameters}.
The total number of antennas is $M = 1111$, the array aperture is $L = 50$\,cm, and the Fraunhofer array distance is
\begin{equation} \label{eq:Fraunhofer_distance}
    d_{FA} = \frac{2L^2}{\lambda} \approx 50\,\text{m}.
\end{equation}

\begin{table}[t]
    \renewcommand{\arraystretch}{1.2}
    \centering
    \begin{tabular}{c|c}
    \textbf{Parameter} & \textbf{Value} \\
    \hline
    Carrier Frequency, $f_c$ & $30\,\, [\mathrm{GHz}]$ \\ \hline
    Wavelength, $\lambda$ & $1$ [cm] \\ \hline
    Bandwidth, $B$ & $100\,[\mathrm{MHz}]$ \\ \hline
    Noise power, $N_0$ & $-166$ [dBm/Hz] \\ \hline
    Number of horizontal antennas, $M_H$ & $101$ \\ \hline
    Number of vertical antennas, $M_V$ & $11$ \\ \hline
    Number of RF chains, $\rm N_{\rm{RF}}$ & $50$ \\ \hline
    Horizontal antenna spacing, $\delta_H$ & $\lambda/2 = 0.5$ [cm] \\ \hline
    Vertical antenna spacing, $\delta_V$ & $\lambda/2 = 0.5$ [cm] \\ \hline
    \end{tabular}
    \caption{System parameters.\vspace{-0.3cm}}
    \label{tab:system_parameters}
\end{table}

\subsubsection{Case A ($N_{\rm{RF}}\tau > M$)} \label{sec:scenario_a}
Assume $\tau = 100$, which yields $N_{\rm{RF}}\tau = 5000 > M$. Fig.~\ref{fig:3a} shows the NMSE of channel estimates versus the distance $R$. We observe that the \ac{LS} estimator is accurate in the near field, where the \ac{SNR} is higher, and degrades in the far field.
We conclude that, when more observations than antennas are available, the near-field channel is accurately estimated using the conventional \ac{LS} estimator, making dictionary-based approaches unnecessary.

For comparison, Fig.~\ref{fig:3a} also reports the \ac{P-SOMP} performance using the far-field dictionary in both the on-dictionary (blue) and off-dictionary (black) scenarios. Since the near-field channel is not sparse in the far-field dictionary domain, \ac{P-SOMP} poorly performs in both cases at short distances. As the distance $R$ increases, the channel becomes sparser, but the \ac{SNR} decreases. In the on-dictionary case, the benefit of increased sparsity outweights the \ac{SNR} degradation, improving NMSE. In contrast, in the off-dictionary case, these two effects balance, yielding an approximately constant NMSE.

\subsubsection{Case B ($N_{\rm{RF}}\tau < M$)} \label{sec:far_field_scenario_b}
\begin{figure}[t]   
    \begin{subfigure}{\columnwidth}
        \centering\vspace{-0.3cm}
        \includegraphics[width=\columnwidth]{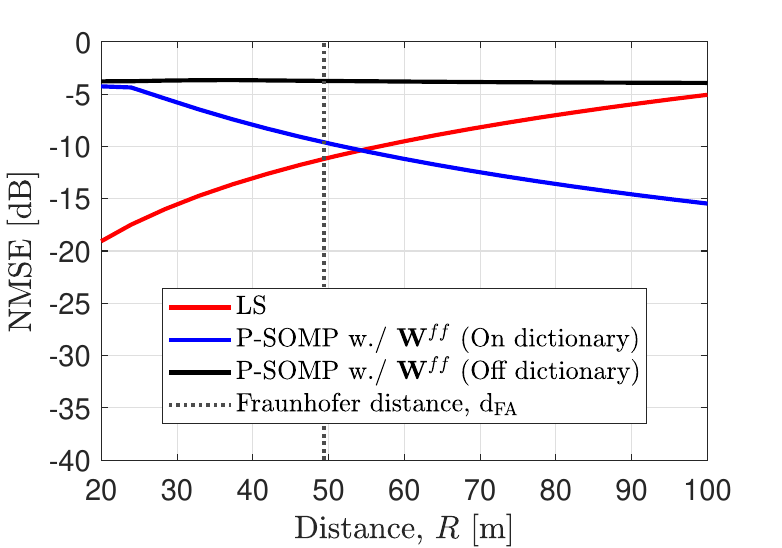}
        \caption{Case A: $N_{\rm{RF}}\tau = 5000 > M$.}
        \label{fig:3a}
    \end{subfigure}
    \begin{subfigure}{\columnwidth}
        \centering
        \includegraphics[width=\columnwidth]{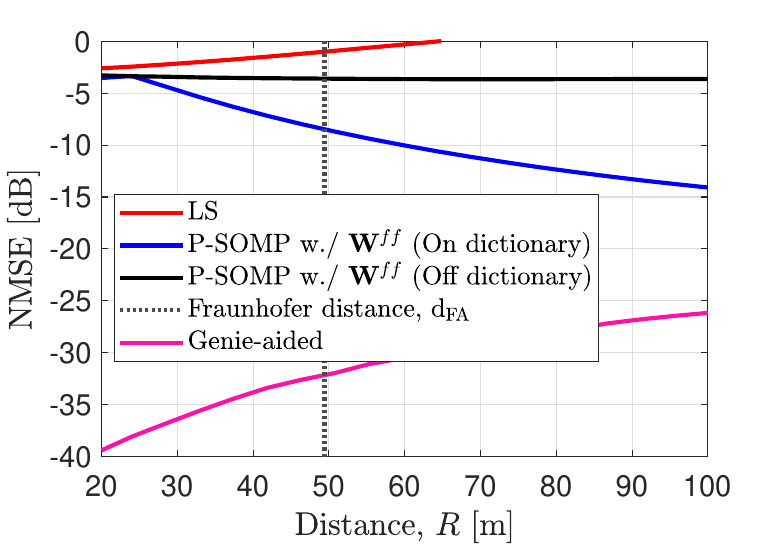}
        \caption{Case B: $N_{\rm{RF}}\tau = 500 < M$.}
        \label{fig:3b}
    \end{subfigure}
    \caption{Average NMSE of channel estimates as a function of the distance $R$. The number of observations is $N_{\rm{RF}}\tau = \{5000,500\}$ in scenarios A and B, respectively.}
\end{figure}

We now consider a scenario with fewer pilot symbols, setting $\tau = 10$ while keeping all other parameters unchanged. This reduces the number of observations to $N_{\rm{RF}}\tau = 500$.
Fig.~\ref{fig:3b} shows the NMSE versus the distance $R$. Since $N_{\rm{RF}}\tau < M$, the \ac{LS} estimator becomes inaccurate regardless of the \ac{SNR}. For comparison, the \ac{P-SOMP} performance using the far-field dictionary is evaluated in the on-dictionary and off-dictionary cases. Since the near-field channel is not sparse in the far-field dictionary domain, both cases yield poor performance at short distances. As $R$ increases, the channel becomes sparser, but the \ac{SNR} decreases. In the on-dictionary case, increased sparsity dominates, improving NMSE, whereas in the off-dictionary case these effects balance, resulting in an approximately constant NMSE.

With $N^\mathcal{P}=1$, channel sparsity significantly limits performance. To illustrate the impact of sparsity, we also consider an ideal \textit{Genie-aided} \ac{UE}-dependent dictionary, which can be chosen as any orthonormal basis of $\mathbb{C}^M$ that includes the true channel. In this case, the channel is maximally sparse, and the corresponding \ac{P-SOMP} performance (purple curve in Fig.~\ref{fig:3b}) significantly outperforms the far-field dictionary, particularly in the near field.

We conclude that, when $N_{\rm{RF}}\tau < M$, the \ac{LS} estimator is inaccurate and the traditional far-field dictionary fails to provide sparse channel representations in the near field, leading to poor estimation accuracy. However, effective channel estimation remains possible using appropriate dictionaries to make the channel sparser. This motivates the development of alternative dictionary designs that are better suited to near-field channel estimation. The proposed one is introduced in the next section.

\section{Proposed near-field grid design}\label{sec:proposed_design}
\begin{figure*}[t]
     \centering
     \begin{subfigure}[b]{\columnwidth}
         \centering\vspace{-0.3cm}
         \includegraphics[width=\textwidth]{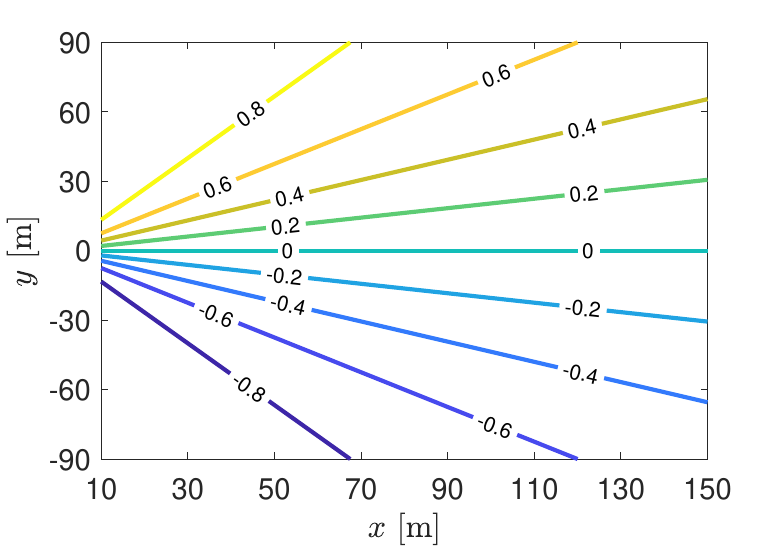}
         \caption{$b = 0$ m}
         \label{fig:4a}
     \end{subfigure}
     \hfill
         \begin{subfigure}[b]{\columnwidth}
         \centering\vspace{-0.3cm}
         \includegraphics[width=\textwidth]{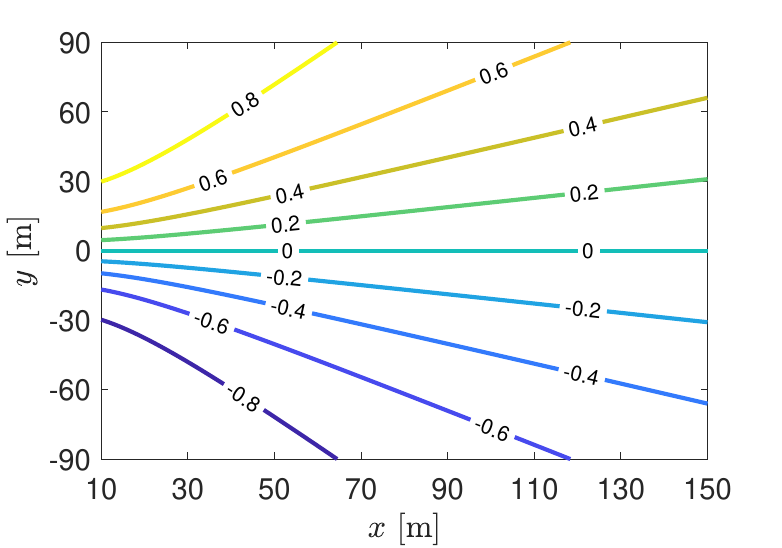}
          \caption{$b = 20$ m}
          \label{fig:4b}
     \end{subfigure}  
     \caption{$\mathcal{L}$-curves $\Gamma(R,\varphi)$ for different constant values and  $b = 0$ or $20$\,m.\vspace{-0.3cm}}
\end{figure*}

Consider two arbitrary points on the \ac{RP}:
\begin{align}
\mathbf{p} &= (\rho_p \cos\varphi_p,\rho_p \sin\varphi_p,-b) \label{eq:p} \\
\mathbf{q} &= (\rho_q \cos\varphi_q, \rho_q \sin\varphi_q,-b) \label{eq:q}
\end{align}
and denote by $R_p$ and $R_q$ their respective distances to the center of the \ac{UPA}. Let
\begin{equation} \label{eq:correlation_1}
    \mu
    = \frac{1}{M}\left|\vect{s}^{\Htran}(\rho_q,\varphi_q,b)\vect{s}(\rho_p,\varphi_p,b)\right|
\end{equation}
be the magnitude of the normalized correlation between the steering vectors associated with $\mathbf{p}$ and $\mathbf{q}$, given by
\begin{align}
    \vect{s}(\rho_p,\varphi_p,b) &= \big[e^{\iu\frac{2\pi}{\lambda}||\vect{p}-\vect{u}_1||},\ldots,e^{\iu\frac{2\pi}{\lambda}||\vect{p}-\vect{u}_M||}\big]^\Ttran \label{eq:steeringvector_p}\\
    \vect{s}(\rho_q,\varphi_q,b) &= \big[e^{\iu\frac{2\pi}{\lambda}||\vect{q}-\vect{u}_1||},\ldots,e^{\iu\frac{2\pi}{\lambda}||\vect{q}-\vect{u}_M||}\big]^\Ttran. \label{eq:steeringvector_q}
\end{align}
In Appendix~\ref{app:B}, it is shown that, if the parabolic approximation holds, then $\mu$ is well-approximated as:
\begin{equation} \label{eq:correlation_2}
    \mu  = \mu_H \left(\Gamma_p, \Gamma_q, R_p, R_q\right) \mu_V \left(R_p, R_q,b\right)
\end{equation}
where
\begin{align}
    \!\!\!\!\!\!\mu_H \left(\Gamma_p, \Gamma_q\right)= \dfrac{1}{M_H} \left |  \sum_{m = - (M_H-1)/2}^{(M_H-1)/2} \!\!\!\!\!\!\!\!\!e^{\iu \left(A_H m +B_Hm^2\right) }\right | \label{eq: mu_H}
\end{align}
and
\begin{align}
    \mu_V \left(R_p, R_q,b\right)= \dfrac{1}{M_V}\left | \sum_{m = - (M_V-1)/2}^{(M_V-1)/2} \!\!\!\!\!\!\!\!\!\!e^{\iu \left(A_Vm  + B_Vm^2\right)}\right | \label{eq: mu_V}
\end{align}
with
\begin{align} \label{eq: AH}
A_H &= \dfrac{2 \pi \delta_H} {\lambda} \left(\Gamma_q -\Gamma_p\right) \\
B_H &= \dfrac{\pi \delta_H^2} {\lambda} \left[ \dfrac{1} {R_p}\left(1 - \Gamma_p^2 \right) - \dfrac{1} {R_q}\left(1 - \Gamma_q^2 \right) \right] \label{eq: BH}
\end{align}
and
\begin{align}
    \label{eq:AV}
    A_V &= \dfrac{2 \pi \delta_V b} {\lambda}  \left( \dfrac{1}{R_p} - \dfrac{1}{R_q}\right) \\
    \label{eq:BV}
    B_V &= \dfrac{\pi \delta_V^2}{\lambda} \left[\dfrac{1}{R_p}\left(1-\dfrac{b^2}{R^2_p}\right) -  \dfrac{1}{R_q}\left(1-\dfrac{b^2}{R^2_q}\right)\right].
\end{align}
Also, we have defined
\begin{equation} \label{eq:Gamma_i}
    \Gamma_i = \sqrt{1 - \dfrac{b^2}{R^2_i}} \sin \varphi_i \quad \quad i = p,q.  
\end{equation}

The following special cases can be retrieved: 

\begin{itemize}
    \item If $R_p$ and $R_q$ are much larger than $b$, then $\Gamma_i = \sin \varphi_i$ for $i=p,q$ and thus the coefficients above become
\begin{align}
    A_H &= \dfrac{2 \pi \delta_H} {\lambda} \left(\sin \varphi_q-\sin \varphi_p\right)
\end{align}
and $A_V = B_V = 0$. Accordingly, \eqref{eq:correlation_2} reduces to the correlation function used in far field.
\item If $b =0$, we have that $\Gamma_i = \sin \varphi_i$ for $i=p,q$, and thus \begin{align} \label{eq: AH_1}
    A_H &= \dfrac{2 \pi \delta_H} {\lambda} \left(\sin \varphi_q-\sin \varphi_p\right) 
\end{align}
and
\begin{align} \label{eq: AV_1}
    A_V &= 0 \\ \label{eq: BV_1}
    B_V &= \dfrac{\pi \delta_V^2}{\lambda} \left(\dfrac{1}{R_p} -  \dfrac{1}{R_q}\right).
\end{align}
From \eqref{eq: AH_1}-\eqref{eq: BV_1}, it easily follows that $\mu$ reduces to the correlation function used in \cite[Eq. (10)]{demir2023new}.

\end{itemize}

The proposed grid design is based on the selection of particular curves and circles on the \ac{RP}, referred to as \textit{$\mathcal{L}$-curves} and \textit{\ac{RP}-circles}, which will be defined in the next two sections, starting from the expressions of the correlation $\mu$ given in \eqref{eq:correlation_2}-\eqref{eq:BV}. The grid points are the intersection points between the selected $\mathcal{L}$-curves and \ac{RP}-circles.

\subsection{Level Curves}
A \textit{level curve} ($\mathcal{L}$-curve) is defined as the set of points on the \ac{RP} for which
\begin{equation} \label{eq:level_curve}
    \Gamma(R,\varphi) = \sqrt{1 - \dfrac{b^2}{R^2}} \sin \varphi = \rm{constant}
\end{equation}
with $|\Gamma(R,\varphi)| \leq 1$. From \eqref{eq:Gamma_i} and \eqref{eq:level_curve}, it follows that points $\vect{p}$ and $\vect{q}$ lie on the $\mathcal{L}$-curves $\Gamma(R,\varphi) = \Gamma_i$, with $i = p,q$. 
Notice that \eqref{eq:level_curve} generalizes the curve
\begin{equation}
    \Phi(\varphi) = \sin \varphi = \text{constant}
\end{equation}
introduced in \cite[Eq.~(11)]{cui2022channel}. In particular, the two curves coincide when the height offset vanishes ($b = 0$), or in the asymptotic case $R \to \infty$ (regardless of $b$). Figs.~\ref{fig:4a} and~\ref{fig:4b} show the $\mathcal{L}$-curves $\Gamma(R,\varphi) = 0, \pm 0.2, \pm 0.4, \pm 0.6, \pm 0.8$, for $b = 0,20$ m. Notably, for $b=0$ the $\mathcal{L}$-curves are straight lines, whereas for $b > 0$, they bend, with the curvature becoming more pronounced as $b$ increases.

\begin{figure}[t]
    \centering\vspace{-0.3cm}
    \includegraphics[width=\columnwidth]{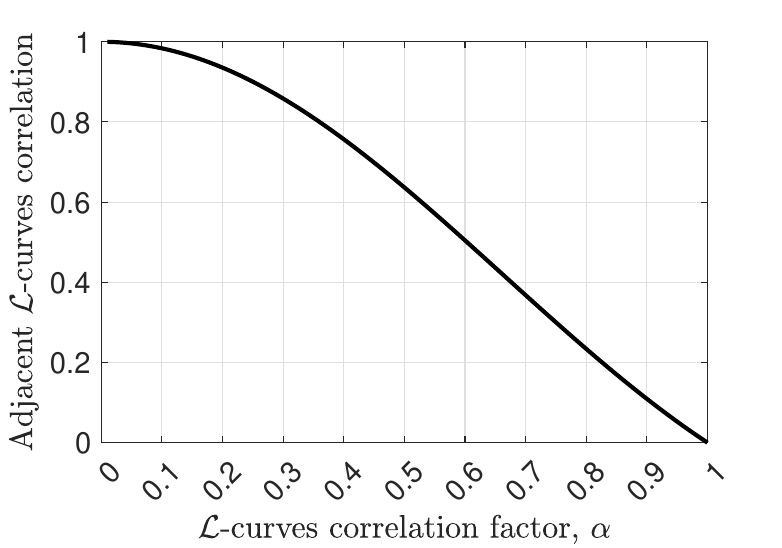}
    \caption{Angular correlation of the steering vectors of two grid points on adjacent $\mathcal{L}$-curves at same distance $R$ from the BS.\vspace{-0.3cm}}
    \label{fig:5}
\end{figure}

Let the points $\vect{p}$ and $\vect{q}$ be located at the same distance $R$ from the array center (i.e., $R_p = R_q = R$) but lie on different $\mathcal{L}$-curves (i.e., $\Gamma_q \neq \Gamma_p$). As shown in Appendix~\ref{app:C}, if $R \gg L_H$, which is typically the case in practice, the correlation $\mu$ simplifies to
\begin{equation} \label{eq:correlation_3}
    \mu = \dfrac{1}{M_H} \left |  \sum_{m = - (M_H-1)/2}^{(M_H-1)/2} e^{\iu \frac{2 \pi \delta_H m} {\lambda} \left(\Gamma_q -\Gamma_p\right)}\right|.
\end{equation}
Consequently, $\mu$  can be controlled by fixing $\Gamma_p-\Gamma_q$. In practice, the first design constraint is the correlation between grid points on \emph{adjacent} $\mathcal{L}$-curves. We select the 
$\mathcal{L}$-curves as
\begin{equation} \label{eq:Gamma_k}
    \Gamma(R,\varphi) = \Gamma^{(k)} = k \alpha \frac{\lambda}{M_H \delta_H}  
\end{equation}
with $0 \le \alpha \le 1$ being a design parameter and
\begin{equation}
    k=-K_{\max},-K_{\max}+1, \dots, K_{\max}-1, K_{\max}
\end{equation}
with
\begin{equation} \label{eq:K_max}
    K_{\max} = \left\lfloor\frac{M_H\delta_H}{\alpha\lambda}\right\rfloor
\end{equation}
begin chosen such that $|\Gamma^{(k)}| \leq 1 $ for all $k$.
If the points $\bf p$ and $\bf q$ lie on adjacent $\mathcal{L}$-curves, then
\begin{equation}
    \Gamma_q -\Gamma_p = \Gamma^{(k+1)} - \Gamma^{(k)} = \alpha \frac{\lambda}{M_H \delta_H}.
\end{equation}
Plugging this result into \eqref{eq:correlation_3} yields 
\begin{align} 
    \mu &= \dfrac{1}{M_H} \left |  \sum_{m = - (M_H-1)/2}^{(M_H-1)/2} e^{\iu \frac{2 \pi m} {M_H}\alpha}\right| = \left|\dfrac{\sin{(\pi \alpha)}}{M_H\sin{\left(\frac{\pi \alpha}{M_H}\right)}}\right| \label{eq:correlation_4}
    \end{align}
which, for $M_H \gg \pi\alpha$ (typically satisfied in practice), becomes
\begin{equation}
    \mu \approx \left| \frac{\sin\left(\pi\alpha\right)} {\pi\alpha}\right|.\label{eq:correlation_5}
\end{equation}
The correlation function becomes independent of both $R$ and $M_H$ and can be directly controlled by the design parameter $\alpha$. The behavior of \eqref{eq:correlation_5} as a function of $\alpha$ is reported in Fig.~\ref{fig:5}. 

\subsection{RP-Circles}

Let $\vect{O}'$ denote the projection of the array center onto the \ac{RP}; see Fig.~\ref{fig:1}. An \ac{RP}-circle of radius $\rho$ is the circle on the \ac{RP} centered at $\vect{O}'$. Let us now assume that the points $\vect{p}$ and $\vect{q}$ lie on the same $\mathcal{L}$-curve (i.e., $\Gamma = \Gamma_p = \Gamma_q$), but are located on different \ac{RP}-circles (i.e., $\rho_p \neq \rho_q$) \footnote{Notice that since$R_i = \sqrt{\rho_i^2+b^2}$ with $i =p,q$, $\rho_p \neq \rho_q$ is equivalent to $R_p \neq R_q$. Without loss of generality, we consider $R_p < R_q$.}. Accordingly, the horizontal correlation $\mu_H$ reduces to
\begin{equation} \label{eq:mu_H_xi}
    \mu_H= \dfrac{1}{M_H} \left |  \sum_{m = - (M_H-1)/2}^{(M_H-1)/2} \!\!\!\!\!\!\!\!\!e^{\iu B_Hm^2 }\right |
\end{equation}
with
\begin{equation} \label{eq:B_H_xi}
    B_H = \dfrac{\pi \delta_H^2 \left(1 - \Gamma^2 \right)} {\lambda}\xi
\end{equation}
where we have defined the design parameter $\xi > 0$ as
\begin{equation}
    \xi = \frac{1}{R_p} - \frac{1}{R_q}.
\end{equation}
Moreover, as shown in Appendix~\ref{app:D}, if
\begin{equation} \label{eq:xi_condition}
    \xi \ll \frac{\lambda}{L_V^2}
\end{equation}
then the coefficients $A_V$ and $B_V$ to compute the vertical correlation $\mu_V$ in \eqref{eq: mu_V}, simplify to
\begin{align} \label{eq:A_V_2}
    A_V &= \dfrac{2 \pi \delta_V b} {\lambda}\xi \\
    B_V &= \dfrac{\pi \delta_V^2}{\lambda} \xi. \label{eq:B_V_xi}
\end{align}

\begin{figure}[t]
    \centering\vspace{-0.3cm}
    \includegraphics[width=\columnwidth]{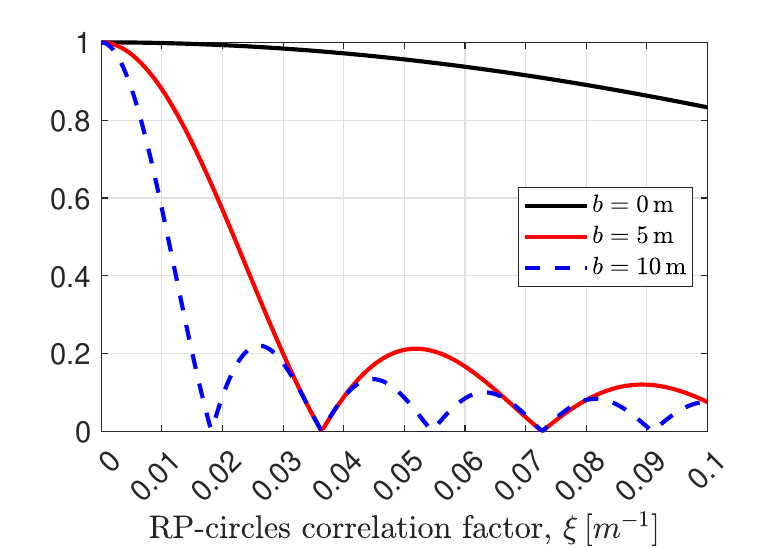}
    \caption{Adjacent \ac{RP}-circles correlation as a function of $\xi$ for different height offsets $b$ and $\mathcal{L}$-curve constant $\Gamma = 0$.\vspace{-0.3cm}}
    \label{fig:6}
\end{figure}

From the above results, it follows that the correlation $\mu$ between points $\vect{p}$ and $\vect{q}$ on \emph{adjacent} \ac{RP}-circle can be controlled through the parameter $\xi$, which therefore provides the second practical design constraint. Accordingly, we select the radii of the \ac{RP}-circles as
\begin{equation} \label{eq: RadiiRP}
    \rho_n = \sqrt{R_n^2 - b^2} \quad \quad n = 0,1,\ldots n^{\ast} -1
\end{equation}
where $R_n$ is computed recursively according to 
\begin{equation} \label{eq:Recursion_A}
    R_n = \dfrac{R_{n-1}}{1 - \xi R_{n-1}} \quad \quad n=1,2,\ldots n^{\ast} -1
\end{equation}
or, equivalently, in closed form as
\begin{equation} \label{eq:Recursion_B}
    R_n = \dfrac{R_{0}}{1 - n \xi R_{0}} \quad \quad n=1,2,\ldots n^{\ast} -1.
\end{equation}
Denoting with $R_{\rm min}$ and $R_{\rm max}$ the minimum and maximum distances from the array center, respectively, we set $R_0 = R_{\min}$ and stop the recursion \eqref{eq:Recursion_A} when either $\xi R_{n-1} \ge 1$ or $R_{n-1} \ge R_{\max}$. This construction guarantees that
\begin{equation}
    \xi = \frac{1}{R_{n-1}}-\frac{1}{R_n} \quad n= 1,2 \dots n^\ast-1
\end{equation}
is independent of $n$. As a result, the design parameter $\xi$ controls the correlation $\mu$ between points lying on the same $\mathcal{L}$-curve and on adjacent \ac{RP}-circles. Since the coefficients $A_V$ in \eqref{eq:A_V_2} and $B_H$ in \eqref{eq:B_H_xi} depend on the height offset $b$ the $\mathcal{L}$-curve constant value $\Gamma$, respectively, the correlation $\mu$ formally depends also on both $b$ and $\Gamma$. However, while the effect of $\Gamma$ is weak and can thus be neglected, the dependence on $b$ is strong and significantly influences the selection of the design parameter $\xi$. The behavior of $\mu$ as a function of $\xi$ is reported in Fig.~\ref{fig:6} for different values of the height offset $b$ and $\mathcal{L}$-curve constant value $\Gamma = 0$. To set the correlation $\mu$ to any arbitrary value in the interval $[0,1]$, we consider, for a given value of $b$, the main lobe of the corresponding curve, which narrows as $b$ increases. Consequently, larger values of $b$ correspond to smaller values of $\xi$, thereby facilitating the satisfaction of \eqref{eq:xi_condition}.

\begin{figure}[t]
     \centering\vspace{-0.3cm}
     \includegraphics[width=\columnwidth]{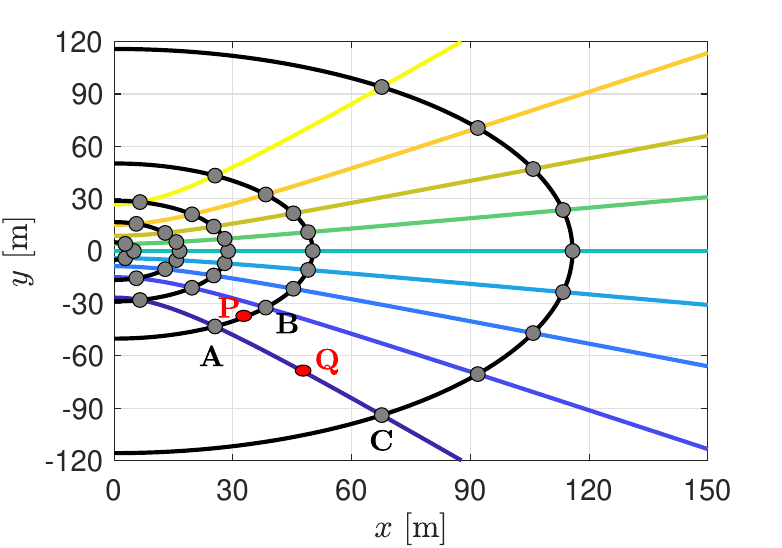}
     \caption{Proposed grid construction by means of $\mathcal{L}$-curves and \ac{RP}-circles for height offset $b = 20$\,m, parameter $\xi = 0.01$ and $\mathcal{L}$-curves selected as $\Gamma^{(k)} = \{-0.8, -0.6 \dots 0.6, 0.8\}.$\vspace{-0.3cm}}
    \label{fig:7}
\end{figure}

\subsection{Grid construction} \label{sec:grid_construction}
As shown in the previous sections, parameters $\alpha$ and $\xi$ can be used to control, respectively, the correlation between points on the same RP-circle and adjacent $\mathcal{L}$-curves, and on the same $\mathcal{L}$-curve and adjacent RP-circles. Accordingly, each value of the pair $(\alpha,\xi)$ corresponds to different sets of $\mathcal{L}$-curves and RP-circles. Once the pair $(\alpha,\xi)$ has been fixed, the grid points are simply given by the intersections between the selected $\mathcal{L}$-curves and \ac{RP}-circles. A pictorial illustration of the grid construction is given in Fig.~\ref{fig:7}, where the small grey disks represent the grid points. For example, $\vect{A}$ and $\vect{B}$ lie on the same RP-circle but belong to different $\mathcal{L}$-curves, whereas $\vect{B}$ and $\vect{C}$ are on the same $\mathcal{L}$-curve but on different RP-circles.

Our grid construction starts with selecting proper values of $\alpha$ and $\xi$. These two parameters can be chosen to control any pair of performance-related metrics, which are relevant to the specific application. We select the values of $\alpha$ and $\xi$ to control the grid size $Q$ and the \textit{optimal NMSE}, defined as
\begin{equation} \label{eq:nmse_opt}
    \text{NMSE}_\text{opt} = 1 - \dfrac{1}{M^2}{\mathbb E}\left\{\underset{\vect{g} \in \vect{G}}{\text{min}} \, |\vect{s}^H(\vect{g})\vect{s}(\vect{r})|^2\right\}
\end{equation}
where $\vect{s}({\vect r})$ and $\vect{s}({\vect g})$ indicate the steering vectors associated with the UE location $\vect{r}$ and with the point $\vect{g} \in \vect{G}$, respectively. Denoting the minimum and maximum distances on the \ac{RP} as 
\begin{align}
    \rho_{\rm min} &= \sqrt{R_{\rm min}-b^2} \\
    \rho_{\rm max} &= \sqrt{R_{\rm max}-b^2}
\end{align}
the expectation in \eqref{eq:nmse_opt} is computed with respect to the {UE} position $\vect{r}$, which is taken as $\vect{r} = \left[\rho\cos{\varphi,\rho \sin{\varphi},-b} \right]$
with $\rho \sim \mathcal{U}(\rho_{\rm min},\rho_{\rm max})$ and $\varphi \sim \mathcal{U}(-\frac{\pi}{2},\frac{\pi}{2})$. Note that $\text{NMSE}_\text{opt}$ can be interpreted as a measure of the average error in approximating the steering vector of an arbitrary point with the steering vector of a nearby grid point\footnote{From this perspective, lower values of $\text{NMSE}_\text{opt}$ are expected to improve the estimation accuracy for grid-based algorithms, such as the \ac{P-SOMP}.}. In this respect, $\text{NMSE}_\text{opt}$ is clearly and strongly related to the values of $\alpha$ and $\xi$. For example, consider a point $\vect P$ located between two adjacent grid points on the same RP-circle (as shown in Fig.~\ref{fig:7}). The error when approximating $\vect{s}(\vect{P})$ with $\vect{s}(\vect{A})$ or $\vect{s}(\vect{B})$ becomes smaller as $\alpha$ decreases, as can be readily deduced from \eqref{eq:correlation_4}. Similarly, consider a point $\vect Q$ lying between two adjacent grid points on the same $\mathcal{L}$-curve (as shown in Fig.~\ref{fig:7}). The error when approximating $\vect{s}(\vect{Q})$ with $\vect{s}(\vect{B})$ or $\vect{s}(\vect{C})$ becomes smaller as $\xi$ decreases. On the other hand, decreasing $\alpha$ and/or $\xi$ increases the grid size. Therefore, selecting the values of $\alpha$ and $\xi$ requires balancing accuracy and complexity. 

\begin{figure}[t]
    \centering\vspace{-0.3cm}
    \includegraphics[width=\columnwidth]{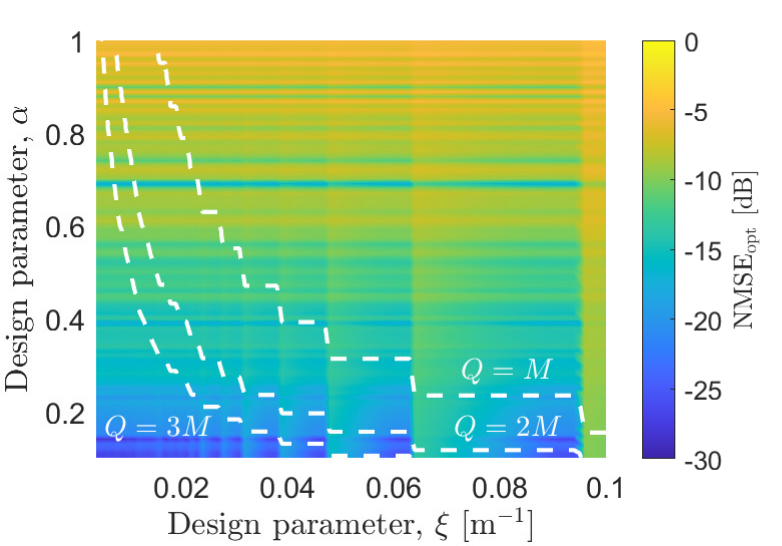}
    \caption{$\rm{NMSE_{opt}}$ as a function of design parameters $\alpha$ and $\xi$.\vspace{-0.3cm}}
    \label{fig:8}
\end{figure}

This trade-off is clearly illustrated in Fig.~\ref{fig:8}, which plots $\text{NMSE}_\text{opt}$ as a function of $\alpha$ and $\xi$. In the same figure, the white curves correspond to the three different values of the grid size $Q=\{M,2M,3M\}$. For any given value of $Q$, the optimal values of $\alpha$ and $\xi$ are those that minimize $\text{NMSE}_\text{opt}$. It is worth noting that \hide{,for $Q > M_H$,} these optimal values typically yield more $\mathcal{L}$-curves\hide{used for grid construction} than \hide{there are} horizontal antennas $M_H$ in the array. In such case, the azimuth angular domain is sampled with more than $M_H$ samples, as it would be adopting the traditional far-field approach. This phenomenon, referred to as \textit{angular over-sampling} \cite{antonelli2026enhancedpolardomaindictionarydesign,ning2026precodingmatrixindicator5g}, suggests that the classical far-field sampling, used in state-of-the-art designs such as \cite{wu2023multiple} and \cite{demir2023new}, is suboptimal in the near field, and should thus be questioned for near-field grid designs.

\begin{algorithm}[t]
\caption{Proposed Grid Construction} 
\label{alg:proposed_grid_construction}
\begin{algorithmic}[1]

\Require Grid size $Q^*$, tolerance $\Delta$, number of samples $N_\xi$, height offset $b$, $\{R_{\rm min}, R_{\rm max}\}$
\Ensure Proposed grid $\vect{G}$

\Statex \textbf{Initialization:}
\State $\vect{\xi} = \left[\frac{\xi_{\rm max}}{N_\xi}, \frac{2\xi_{\rm max}}{N_\xi}, \dots, \xi_{\rm max}\right] \in \mathbb{R}^{N_\xi}$ with $\xi_{\rm max} = \frac{\lambda}{10 L_V^2}$
\State $\{\alpha_{\rm min}, \alpha_{\rm max}\} = \{0,1\}$
\State Initialize $Q^* = 0$ and $\rm NMSE_{opt}^* = 1$

\Statex \textbf{Iteration:}
\For{$i = 0$ \textbf{to} $N_\xi-1$}
    \State Compute $R_n$ according to \eqref{eq:Recursion_A} or \eqref{eq:Recursion_B} with $\xi = \xi_{i+1}$
    \Repeat
        \State $\alpha \gets \frac{\alpha_{\rm min} + \alpha_{\rm max}}{2}$
        \State Compute $K_{\rm max}$ according to \eqref{eq:K_max}
        \State Compute $\Gamma_k$ according to \eqref{eq:Gamma_k}
        \State Construct $\vect{G}$ as the set of intersections between the selected $\mathcal{L}$-curves and \ac{RP}-circles
        \State $Q^* \gets$ number of columns of $\vect{G}$
        \If{$Q^* < Q-\Delta$}
            \State $\alpha_{\rm max} \gets \alpha$
        \Else
            \State $\alpha_{\rm min} \gets \alpha$
        \EndIf
    \Until{$Q-\Delta \le Q^* \le Q+\Delta$}
    
    \State Compute the dictionary associated with the grid $\vect{G}$
    \State Compute $\rm NMSE_{opt}$ according to \eqref{eq:nmse_opt}
    \If{$\rm NMSE_{opt} < NMSE_{opt}^*$}
        \State $\vect{G}^* \gets \vect{G}$
        \State $\rm NMSE_{opt}^* \gets \rm NMSE_{opt}$
    \EndIf
\EndFor

\State \Return $\vect{G}^*$

\end{algorithmic}
\end{algorithm}

\subsection{Summary of grid construction}
The proposed grid construction can be summarized in the following three steps:
\begin{enumerate}
    \item Determine $\alpha$ and $\xi$ such that the grid size equals $Q = M$ and $\rm{NMSE}_{opt}$ is minimized.
    \item Substitute $\alpha$ in \eqref{eq:Gamma_k} to compute the level curves constants $\Gamma_k$, and the value of $\xi$ into \eqref{eq:Recursion_A} or, equivalently, \eqref{eq:Recursion_B} to determine the radii of \ac{RP}-circles.
    \item Obtain the grid points as the intersections of the selected $\mathcal{L}$-curves and \ac{RP}-circles.
\end{enumerate}

\begin{figure}[t]
    \centering\vspace{-0.5cm}
    \includegraphics[width=\columnwidth]{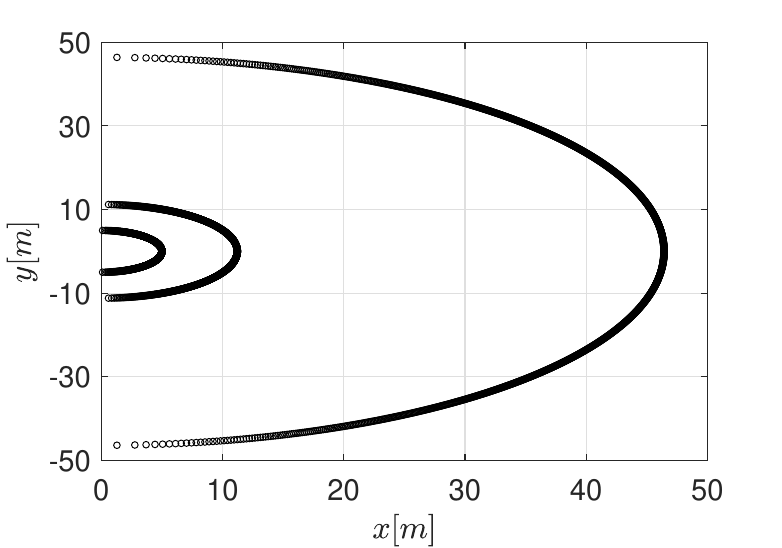}
    \caption{Proposed grid with $\{R_{\min},R_{\max}\} = \{7\,\text{m},100\,\text{m}\}$ and $\{\alpha,\xi\} = \{0.07,0.06\}$.\vspace{-0.3cm}}
    \label{fig:9}
\end{figure}

The pseudo-code for grid construction is given in \textbf{Algorithm~\ref{alg:proposed_grid_construction}}. Additionally, the MATLAB code can be provided upon request and will be made freely available after completion of the revision process. An example of the proposed grid is illustrated in Fig.~\ref{fig:9}. Since the number of level curves is $1443 > M = 1111$, the grid is angularly over-sampled.

\section{Proposed grid performance evaluation} \label{sec:proposed_grid_performance}
We now study the performance of the proposed grid in terms of channel estimation accuracy and SE using the \ac{P-SOMP}.

Assume that $K = 10$ active \acp{UE} are randomly distributed within a volume of interest, shaped as a prism with an angular sector base and centered on the \ac{RP}, see Fig.~\ref{fig:10}.

Specifically, let $\{\rho_{\min},\rho_{\max}\}$ denote the minimum and maximum distances from $\vect{O}'$, $\{\varphi_{\min},\varphi_{\max}\}$ the minimum and maximum azimuth angles, and let $\epsilon$ indicate the height (or thickness) of the prism. The volume of interest consists of all points with coordinates $(\rho\cos{\varphi},\rho\sin{\varphi},z)$ with $\rho \in [\rho_{\min},\rho_{\max}]$, $\varphi \in [\varphi_{\min},\varphi_{\max}]$ and $z \in \left[-b -\epsilon/2,-b + \epsilon/2\right]$. We consider $\{\rho_{\min},\rho_{\max}\} = \{5 \rm{m}, 100\rm{m}\}$ and $\{\varphi_{\min},\varphi_{\max}\} = \{-\pi/3,\pi/3\}$.

Additionally, we assume that each \ac{UE} transmits with the same power $p$, which is set to control the SNR of the signal received from the maximum distance $R_{\max} = \sqrt{\rho_{\max}^2+b^2}$, denoted as $\eta$. Unless otherwise specified, we assume $\eta = 0$\,dB. We set the size of the proposed grid to $Q = M$.
All other simulation parameters are the same as in the scenario $B$, described in Sec.~\ref{sec:far_field_scenario_b}, and are reported in Tab.~\ref{tab:system_parameters}.

\subsection{2D scenario ($\epsilon = 0$)} \label{sec:2D_scenario}
\begin{figure}[t]
    \centering\vspace{-0.3cm}
    \includegraphics[width=0.9\columnwidth]{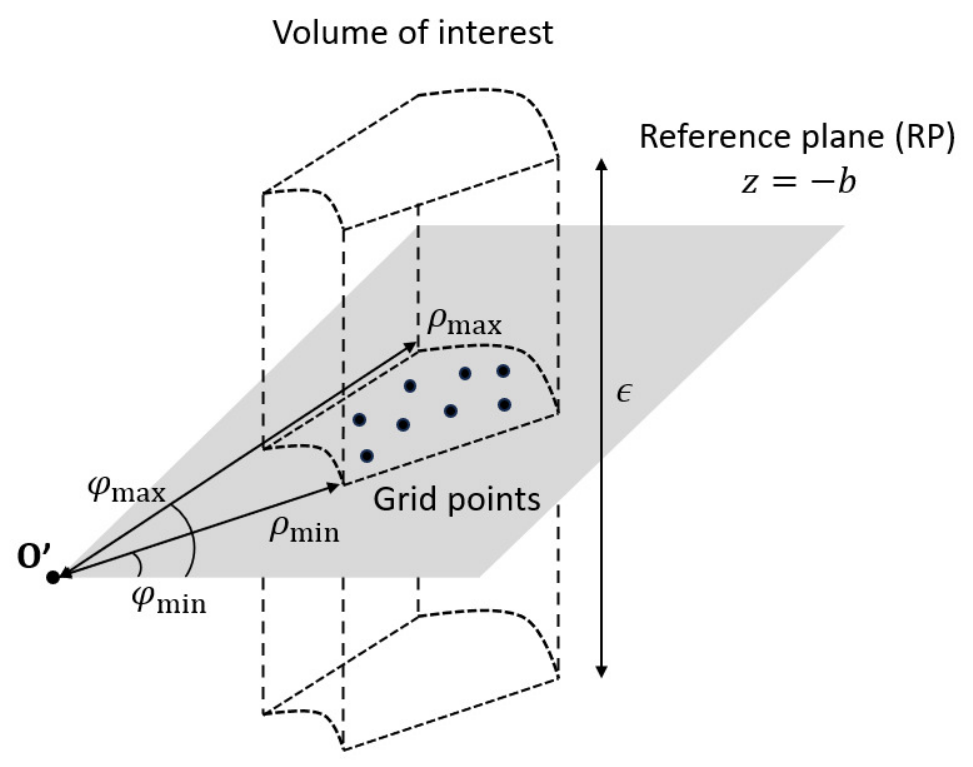}
    \caption{Pictorial illustration of the volume of interest. The proposed grid lies on the \ac{RP}.\vspace{-0.3cm}}
    \label{fig:10}
\end{figure}

First, we assume that \acp{UE} lie on the angular sector on the \ac{RP}, shown in Fig~\ref{fig:10}, i.e., we address the case $\epsilon = 0$. For comparison, we consider a polar-uniform grid on the \ac{RP} with the same size $Q$ as the proposed grid, in which both angular and distance domains are uniformly sampled. Specifically, the polar-uniform grid points are defined as $(\rho_n\cos{\varphi_m},\rho_n \sin{\varphi_m},-b)$ with
\begin{equation}
    \rho_n = \rho_{\min} + \frac{\rho_{\max}-\rho_{\min}}{N_\rho-1}n \quad n = 0,1,\dots N_\rho-1
\end{equation}
and
\begin{equation}
    \varphi_m = \arcsin{\left(\sin{\varphi_{\min}} + \frac{\sin{\varphi_{\max}}-\sin{\varphi_{\min}}}{N_\varphi-1}m\right)}
\end{equation}
where $m = 0,1,\dots N_\varphi-1$. Similarly to \cite{cui2022channel,wu2023multiple,demir2023new}, we consider as many samples in the azimuth domain as the number of horizontal antenna, i.e., $N_\varphi = M_H$. Then, we select the number of  samples in the distance domain $N_\rho$ so that the grid size is equal to $Q$, i.e., $N_\rho = \lfloor Q/N_\varphi\rfloor$.

Moreover, when $b \neq 0$, we also consider a mismatched grid of size $Q$, which is constructed according to the proposed design assuming $b = 0$, regardless of the true value of $b$.
\begin{figure}[t]   
    \centering
    \includegraphics[width=.45\textwidth]{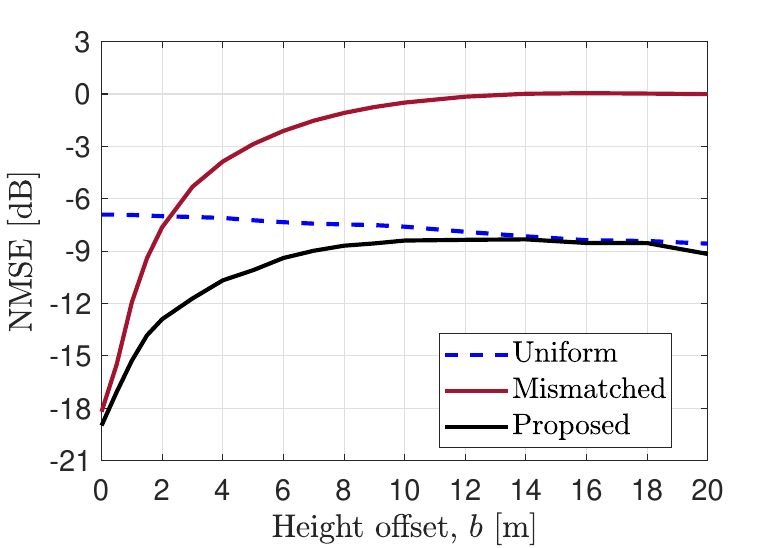}
    \caption{Average NMSE of channel estimates using the \ac{P-SOMP} with various grid designs as a function of the height offset $b$.\vspace{-0.3cm}}
    \label{fig:11}
\end{figure}
\begin{figure}[t]
    \begin{subfigure}[b]{\columnwidth}
         \centering
         \includegraphics[width=\textwidth]{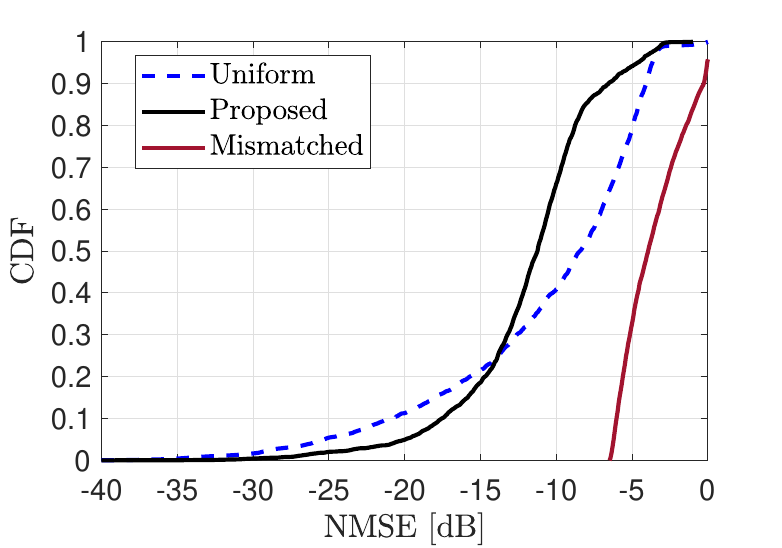}
          \caption{NMSE.\vspace{-0.1cm}}
         \label{fig:12a}
     \end{subfigure}
     \begin{subfigure}[b]{\columnwidth}
     \includegraphics[width=\columnwidth]{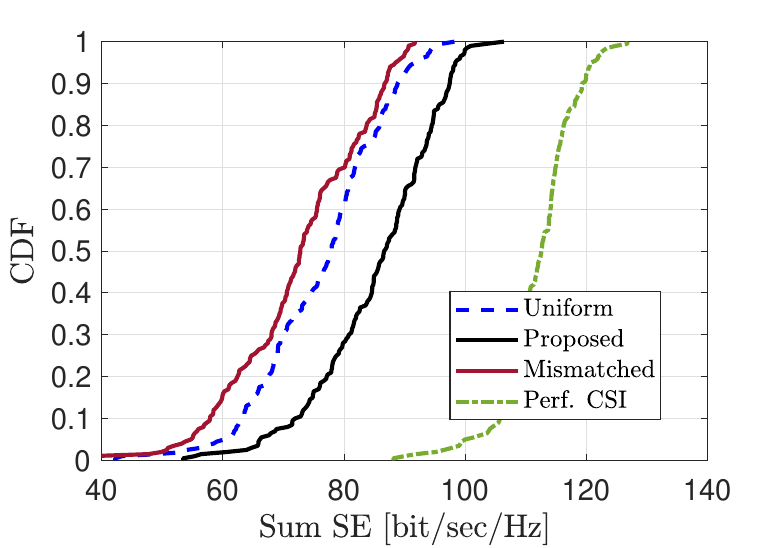}
    \caption{SE.}
    \label{fig:12b}
    \end{subfigure}
    \caption{CDF of NMSE channel estimates and of SE using the \ac{P-SOMP} algorithm with various grid designs in the $2$D scenario ($\epsilon = 0$) for $b = 5$\,m.\vspace{-0.5cm}}
\end{figure}

We begin our analysis of the proposed grid performance by investigating the impact of the height offset $b$. Controlling the SNR by setting $\eta = 0$\,dB, Fig.~\ref{fig:11} shows the average NMSE of channel estimates as a function of $b$. First, we observe that the mismatched grid performance rapidly decreases as $b$ increases, rendering it unviable when $b$ is larger than a few meters. Second, the proposed grid significantly outperforms the uniform-polar design up to approximately $b = 5$\,m, while their performance becomes progressively similar for larger values of $b$. This behavior is further illustrated in Fig.~\ref{fig:12a}, which shows the CDF of NMSE of channel estimates for $b = 5$\,m. The polar-uniform grid yields more very accurate channel estimates (i.e., low NMSE values) as well as more very poor estimates (i.e., large NMSE values) than the proposed grid. As a result, the two designs exhibit comparable accuracy on average. However, as shown in Fig.~\ref{fig:12b}, since SE is more influenced by the reduced number of poor channel estimates, the proposed design achieves higher SE.

\begin{figure}[t]   
    \centering\vspace{-0.3cm}
    \includegraphics[width=\columnwidth]{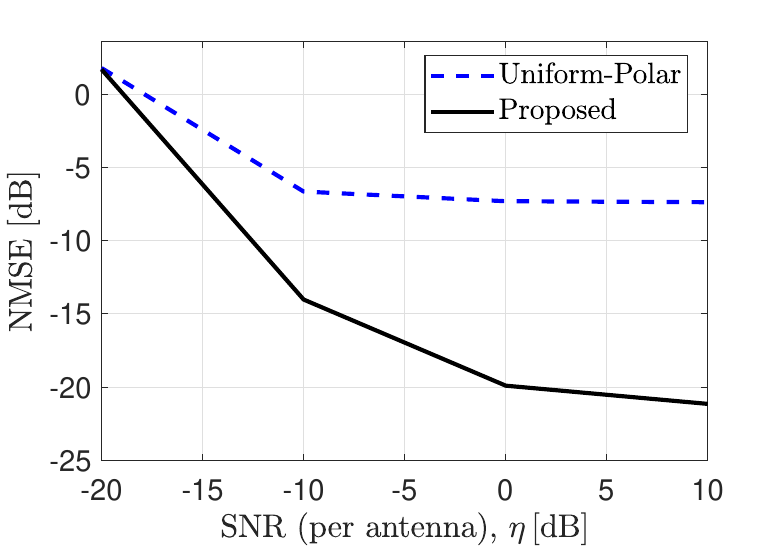}
    \caption{Average NMSE of channel estimates using the \ac{P-SOMP} algorithm with various grids as a function of $\eta$ in the $2$D scenario ($\epsilon = 0$) for $b = 0$.\vspace{-0.3cm}}
    \label{fig:13}
\end{figure}

To conclude the analysis of the $2$D scenario, let us now investigate the impact on channel estimation accuracy of the SNR, which we control through the parameter $\eta$. To this end, Fig.~\ref{fig:13} illustrates the NMSE of channel estimates as a function of $\eta$ with $b = 0$. Notably, as $\eta$ increases, the NMSE converges to a limiting value, which depends on both the grid size and the grid design. In general, such value should be as low as possible to obtain more accurate channel estimates. Moreover, it is desirable that the convergence occurs at higher SNR values, since operating in the non-saturated regime allows further SNR increases to improve channel estimation accuracy. From Fig.~\ref{fig:13}, we observe that the proposed design both converges to a significantly lower value than the uniform-polar and does so at a higher SNR, supporting the proposed design validity.

\subsection{3D scenario $(\epsilon > 0)$} \label{sec:3D_scenario}
The proposed design locates grid points on a single plane, namely the \ac{RP}. While these points are used for channel estimation, \acp{UE} are not necessarily displaced on a planar surface in practice, but may be distributed within a three-dimensional volume instead. To discuss the robustness of the proposed grid under such conditions, we now address the case $\epsilon > 0$.

To this end, we evaluate the impact of the thickness $\epsilon$ on both channel estimation accuracy and SE using the proposed grid design. For comparison, we consider the two state-of-the art near-field grid designs in \cite{bjornson2021primer} and \cite{wu2023multiple}, from Demir-Bjornson and Wu-Dai, respectively. While such designs assume that \acp{UE} are distributed across the entire half-space $x \geq 0$, we consider a limited volume of interest in our scenario. To ensure a fair comparison, we build all considered grids so that they share the same number of grid points falling within this volume, which we set to $M$, and discard all grid points outside it. Additionally, we also consider the traditional far-field dictionary reviewed in Sec.~\ref{sec:far_field_review}. 
and assume that the height offset is $b = 0$.  

\begin{figure}[t]   
    \centering
    \includegraphics[width=\columnwidth]{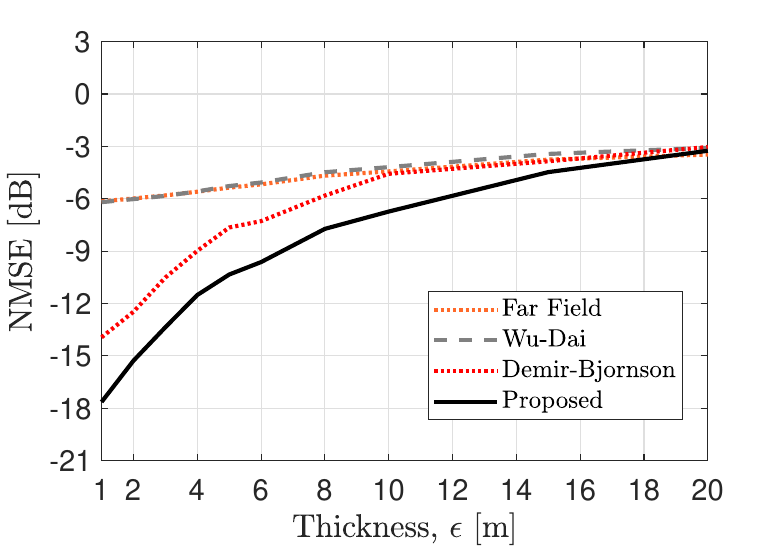}
    \caption{Average NMSE of channel estimates using the \ac{P-SOMP} with various grid designs as a function of the thickness $\epsilon$ for $b = 0$.\vspace{-0.3cm}}
    \label{fig:14}
\end{figure}
Fig.~\ref{fig:14} shows the NMSE of channel estimates as a function of the thickness $\epsilon$. Notably, for values of $\epsilon$ up to approximately $10$\,m, the proposed grid significantly outperforms all other considered designs, while, for larger values of $\epsilon$, the performance of all designs progressively becomes comparable. 

To conclude the $3$D scenario performance analysis, we consider a final setup that commonly occurs in practice. Specifically, we assume that \acp{UE} are randomly distributed at heights between $0$ and $10$\,m, while the \ac{BS} is positioned at a height of $5$\,m. Accordingly, we set $\epsilon = 10$\,m and $b = 0$\,m, while keeping all other simulations parameters unchanged.

\begin{figure}[t]   
    \begin{subfigure}{\columnwidth}
        \centering\vspace{-0.3cm}
        \includegraphics[width=\columnwidth]{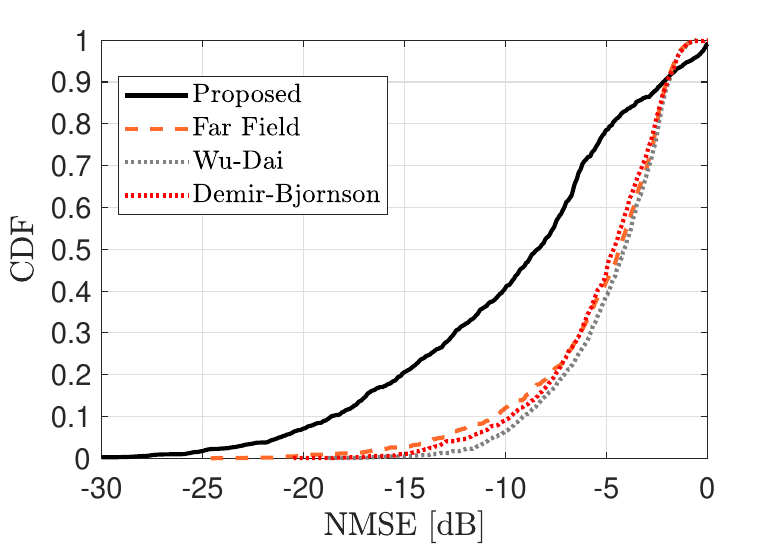}
        \caption{NMSE.}
        \label{fig:15a}
    \end{subfigure}
    \hfill
    \begin{subfigure}{\columnwidth}
        \centering
            \includegraphics[width=\columnwidth]{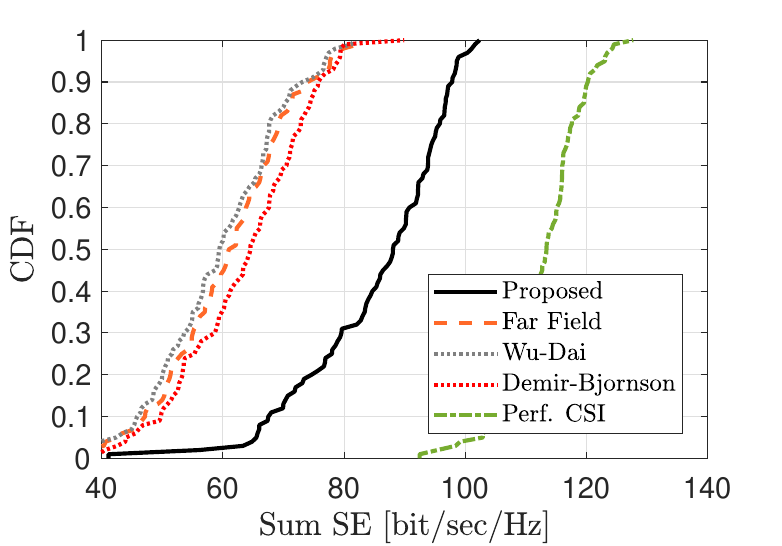}
        \caption{Sum SE.}
        \label{fig:15b}
    \end{subfigure}
    \caption{CDF of NMSE of channel estimates and of aggregate SE using the \ac{P-SOMP} algorithm with various grid designs for $\epsilon = 10$\,m and $b =0$.\vspace{-0.3cm}}
\end{figure}
Fig.~\ref{fig:15a} and~\ref{fig:15b} show the CDFs of NMSE of channel estimates and of SE, respectively. Remarkably, the proposed grid outperforms all other designs -- which exhibit comparable performance -- in both metrics. Moreover, a notable SE gap remains between the proposed design and the benchmark case with perfect channel knowledge, which motivates further research towards optimal near-field dictionary designs.

\section{Conclusions}
With limited observations, the \ac{LS} is inaccurate, and channel estimation relies on channel sparsity in a suitable dictionary domain. Since near-field channels are not sparse in the traditional far-field dictionary, dedicated near-field designs are typically preferred for near-field estimation. However, existing dictionaries generally provide coverage of the full three-dimensional environment, which is unnecessary when \acp{UE} are located on the ground. Therefore, we proposed a novel near-field grid design tailored to this common scenario, displacing grid points on a reference plane located at an arbitrary height offset with respect to the \ac{BS}. Results showed that, as long as \acp{UE} are not excessively far from the reference plane, the proposed design outperforms state-of-the art approaches in terms of both channel estimation accuracy and spectral efficiency. Furthermore, we used an NMSE metric, whose minimization enhances codebook performance, and remarks that classical far-field angular sampling -- still adopted in recent near-field dictionaries -- is questionable in the near field.

Despite the encouraging results, further research on dictionary-based channel estimation is required. Particularly, the  robustness to multipath \ac{NLOS} scenarios, as well as the impact of directive antennas, should be systematically investigated to enable more realistic performance evaluations. Moreover, state-of-the-art near-field dictionary designs are typically coherence-based; to further increase system performance, future designs should be instead directly based on performance-related metrics, such as the $\rm{NMSE_{opt}}$.

\appendices

\section{} \label{app:B}
In this appendix, we show that if the parabolic approximation is satisfied, then the correlation $\mu$ in \eqref{eq:correlation_1} reduces to \eqref{eq:correlation_2}. To do so, we start plugging into \eqref{eq:correlation_1} the expressions of the steering vectors of $\vect{p}$ and $\vect{q}$ from \eqref{eq:steeringvector_p} and \eqref{eq:steeringvector_q}. This yields
\begin{align} 
    \mu &= \frac{1}{M}\sum_{m=1}^Me^{j\frac{2\pi}{\lambda}\left(||\vect{p} - \vect{u}_m||-||\vect{q}-\vect{u}_m||\right)} \nonumber \\
    &= \frac{1}{M}\sum_{m = 1}^Me^{j\frac{2\pi}{\lambda}\left[r_m^{(p)}-r_m^{(q)}\right]}\label{eq:correlation6}
\end{align}
where $r_m^{(p)}= ||\vect{p} - \vect{u}_m||$ and $r_m^{(q)}= ||\vect{q} - \vect{u}_m||$ denote the distances from the $m$th antenna of the \ac{UPA} to the points $\vect{p}$ and $\vect{q}$, respectively. The following lemma is then used. 

\begin{lemma}[Fresnel approximation] \label{lemma:fresnel_approximation}
Consider an arbitrary point on the \ac{RP} given by $\vect{x} = (\rho\cos{\varphi},\rho\sin{\varphi},-b)$ and denote with $R$ the distance from the point $\vect{x}$ to the center of the \ac{UPA}. Then, define the \textit{Fresnel distance} \cite{selvan2017fraunhofer} as $
R_F= 0.5\sqrt{\frac{L^3}{\lambda}}$.
As shown in \cite{cui2022channel} and~\cite{selvan2017fraunhofer}, if $R \geq R_F$, then the distance $r_m$ from the point $\mathbf{x}$, to the $m$th antenna of the array can be approximated as
\begin{align} \nonumber
        r_m \approx  R & - \dfrac{1}{R} \left[\delta_H \rho \sin\varphi \,i(m)- \delta_V b \,j(m)   \right]\\\nonumber
        & + \dfrac{1}{2R} \Big[\delta^2_H \left(1 - \dfrac{\rho^2}{R^2} \sin^2\varphi\right) i^2(m) \\
        & +  \delta^2_V  \left(1 -\dfrac{b^2}{R^2} \right) j^2(m)\Big]
        \label{eq:ParabolicAppxUPA}
\end{align}
which is commonly referred to as the \textit{Fresnel (or parabolic) approximation} \cite{10070575}.
\end{lemma}

Based on \textbf{Lemma~\ref{lemma:fresnel_approximation}}, if both $R_p \geq R_F$ and $R_q \geq R_F$, then we can approximate the difference $r^{(p)}_m - r^{(q)}_m$ as 
\begin{align}
        r^{(p)}_m - r^{(q)}_m &\approx  \ R_p - R_q \nonumber \\
        & - \delta_H (\Gamma_p - \Gamma_q) i(m) + \delta_V b \left( \dfrac{1}{R_p} - \dfrac{1}{R_q}\right)j(m) \nonumber \\
        & + \dfrac{\delta_H^2}{2} \left[\dfrac{1}{R_p}(1-\Gamma^2_p) -  \dfrac{1}{R_q}(1-\Gamma^2_q)\right] i^2(m) \nonumber \\
        & + \dfrac{\delta_V^2}{2} \left[\dfrac{1}{R_p}\left(1-\dfrac{b^2}{R^2_p}\right) -  \dfrac{1}{R_q}\left(1-\dfrac{b^2}{R^2_q}\right)\right] j^2(m)
        \label{eq:ParabolicAppxUPA_diff}
\end{align}
where we have defined $\Gamma_i$ with $i = p,q$ according to \eqref{eq:Gamma_i}. Plugging \eqref{eq:ParabolicAppxUPA_diff} into \eqref{eq:correlation6} yields $ \eqref{eq:correlation_2}$.


\section{} \label{app:C}
In this appendix, we show that if $R \gg L_H$, then the correlation between the steering vectors of two points lying on adjacent $\mathcal{L}$-curves at the same distance $R$ simplifies to \eqref{eq:correlation_3}. We start assuming that the points $\vect{p}$ and $\vect{q}$ are located at the same distance $R$ from the center of the array (i.e., $R_p = R_q = R$), but lie on adjacent $\mathcal{L}$-curves (i.e., $\Gamma_q = \Gamma^{(k')} \neq \Gamma_p = \Gamma^{(k'')}$ with $k'-k''=1$). As a result, $A_V = B_V = 0$, which yields $\mu_V = 1$. Accordingly, the correlation $\mu$ in \eqref{eq:correlation_1} reduces to 
\begin{align}
    \mu = \mu_H =\left | \dfrac{1}{M_H} \sum_{m = - (M_H-1)/2}^{(M_H-1)/2} e^{\iu (A_H m + B_H m^2)}\right| \label{eq:correlation_7}
\end{align}
where $A_H$ is given in \eqref{eq: AH} and $B_H$ in \eqref{eq: BH} simplifies to
\begin{align}
    \!\!\!\!\!B_H &= \frac{\pi \delta_H^2} {\lambda R} \left(\Gamma_q^2 - \Gamma_q^2 \right)  = \frac{\pi \delta_H^2} {\lambda R} \left(\Gamma_q - \Gamma_p \right)\left(\Gamma_q + \Gamma_p \right) \nonumber \\
    &= \frac{\pi \delta_H^2} {\lambda R} \alpha \frac{\lambda}{M_H\delta_H}\left(\Gamma_p + \Gamma_q \right) =\alpha\frac{\pi \delta_H } {M_HR} \left(\Gamma_p + \Gamma_q \right) .\!\!\label{eq:BH_2}
\end{align}
Based on \eqref{eq:BH_2}, the maximum phase rotation due to the quadratic term $B_H m^2$ can be upper bounded as
\begin{align}
    \max\{|B_H| m^2\} &= |B_H| \dfrac{(M_H-1)^2}{4} \\
    &= \alpha \dfrac{\pi \delta_H}{M_H R} |\Gamma_p + \Gamma_q| \dfrac{(M_H-1)^2}{4} \\
    &= \alpha \dfrac{\pi}{4} \dfrac{M_H-1}{M_H} \dfrac{L_H}{R} |\Gamma_p + \Gamma_q| \\
    &\overset{(a)}{<} \alpha \dfrac{\pi}{2} \dfrac{L_H}{R}
\end{align}
where $\overset{(a)}{<}$ follows from $|\Gamma_p+\Gamma_q|\le 2$, which can be immediately deduced from \eqref{eq:Gamma_i}. Since $0 \leq \alpha \leq 1$, if $R \gg L_H$, the quadratic term $B_H m^2$ can be neglected, which means that the correlation $\mu$ reduces to
\begin{align}
    \mu &\approx \left | \dfrac{1}{M_H} \sum_{m = - (M_H-1)/2}^{(M_H-1)/2} e^{\iu A_H m}\right| \nonumber \\
        &= \dfrac{1}{M_H} \left |  \sum_{m = - (M_H-1)/2}^{(M_H-1)/2} e^{\iu \frac{2 \pi \delta_H m} {\lambda} \left(\Gamma_q -\Gamma_p\right)}\right|
\end{align}
which coincides with \eqref{eq:correlation_3}.


\section{} \label{app:D}
In this appendix, we show that if $\xi \ll \lambda/L_V^2$, then the coefficient $B_V$ can be approximated as in \eqref{eq:B_V_xi}.
Let us start from the general expression of $B_V$, given in \eqref{eq:BV}:
\begin{align}
    B_V &= \dfrac{\pi \delta_V^2}{\lambda} \left[\dfrac{1}{R_p}\left(1-\dfrac{b^2}{R^2_p}\right) -  \dfrac{1}{R_q}\left(1-\dfrac{b^2}{R^2_q}\right)\right] \nonumber \\
    &=\dfrac{\pi \delta_V^2}{\lambda} \left[\xi + b^2\left({\dfrac{1}{R_q^3}-\dfrac{1}{R_p^3}}\right)\right]. \label{eq:BV_2}
\end{align}
Now, we can observe that, if the term $b^2\left({\frac{1}{R_q^3}-\frac{1}{R_p^3}}\right)$ is neglected, the expressions of $B_V$ in \eqref{eq:BV_2} and \eqref{eq:B_V_xi} coincide. Hence, we need to understand when such term is negligible.To this end, let us assume that the arbitrary points $\vect{p}$ and $\vect{q}$ are located on adjacent \ac{RP}-circles (i.e., $R_p = R_{n-1} \neq R_q = R_{n}$ for some $n$). Accordingly, the coefficients of the vertical correlation
\begin{align}
    \mu_V= \dfrac{1}{M_V}\left | \sum_{m = - (M_V-1)/2}^{(M_V-1)/2} \!\!\!\!\!\!\!\!\!\!e^{\iu \left(A_Vm  + B_Vm^2\right)}\right| \nonumber
\end{align}
simplify as 
\begin{align}A_V &= \dfrac{2 \pi \delta_V b} {\lambda}\xi \\
B_V &= \dfrac{\pi \delta_V^2}{\lambda} \left[\xi + b^2\left({\dfrac{1}{R_n^3}-\dfrac{1}{R_{n-1}^3}}\right)\right].
\end{align}
Thus, the maximum phase rotation $\Delta{\theta}=\max\{|B_V| m^2\}$, due to the term $b^2(\frac{1}{R^3_n}-\frac{1}{R^3_{n-1}} )$, is given by
\begin{align}
    \Delta{\theta} &= |B_V|\dfrac{(M_V-1)^2}{4} \\
    &=\left|\dfrac{\pi \delta_V^2}{\lambda} b^2\left(\dfrac{1}{R^3_n}-\dfrac{1}{R^3_{n-1}} \right)\right| \dfrac{(M_V-1)^2}{4}  \\
    &= \dfrac{\pi L_V^2}{4\lambda} b^2\left|\dfrac{1}{R^3_n}-\dfrac{1}{R^3_{n-1}} \right|.
\end{align}
After some algebra, it can be shown that $\Delta\theta < \frac{3 \pi}{4} \frac{L^2_V \xi}{\lambda}$. Accordingly, we conclude that if $\dfrac{L^2_V \xi}{\lambda} \ll 1$ or, equivalently, $\xi \ll \frac{\lambda}{L^2_V}$ the term $b^2(\frac{1}{R^3_n}-\frac{1}{R^3_{n-1}} )$ can be neglected. Consequently, the coefficient $B_V$ simplifies as \eqref{eq:B_V_xi}.

\bibliographystyle{IEEEtran}
\bibliography{refs}

\end{document}